\documentclass[balance=false,sigconf,screen,authorversion,nonacm]{acmart}
\AtBeginDocument{%
  }

\usepackage[ruled,vlined,linesnumbered]{algorithm2e}
\usepackage[sets,operations,colors]{cora-macs}

\newcommand{\interval}[1]{\opBoxEnclosure{#1}}
\newcommand{\reduce}[3]{\operatortt{reduce}{#1,\texttt{\small#2},#3}}
\newcommand{\wrap}{\mathrm{W}}
\newcommand{\boxcost}{\gamma^{\square}}
\newcommand{\abscost}{\gamma^{\triangle}}
\newcommand{\aprxboxcost}{\hat{\gamma}^{\square}}

\newcommand{\abscostcum}{\bar\eta}

\newcommand{\abs}[1]{\lvert #1\rvert}
\newcommand{\norm}[1]{\lVert #1\rVert}

\newcommand{\rad}{\operatorname{rad}}
\newcommand{\mth}[1]{\texttt{#1}}
\newcommand{\girard}{\mth{Girard}}
\newcommand{\combastel}{\mth{Combastel}}
\newcommand{\scott}{\mth{Scott}}
\newcommand{\pca}{\mth{PCA}}
\newcommand{\method}{\mth{RAZeR}}

\usepackage{nicefrac}

\usepackage{stfloats}

\usepackage{cleveref}
\usetikzlibrary{arrows.meta}

\crefname{section}{Sec.}{Sec.}
\crefname{subsection}{Sec.}{Sec.}
\crefname{figure}{Fig.}{Fig.}
\crefname{algorithm}{Alg.}{Alg.}
\crefname{table}{Tab.}{Tab.}
\crefname{example}{Ex.}{Ex.}
\crefname{definition}{Def.}{Def.}
\crefname{proposition}{Prop.}{Prop.}
\crefname{problem}{Problem}{Problem}
\crefname{corollary}{Cor.}{Cor.}
\crefname{theorem}{Thm.}{Thm.}
\crefname{lemma}{Lemma}{Lemmas}
\crefname{appendix}{Appendix}{Appendix}

\renewcommand{\Cref}[1]{\cref{#1}}

\newcommand{\Zorder}{\rho}

\begin{document}

\title{RAZeR: Representation Aware Zonotope Reduction}

\author{Vladimir Krsmanovi\' c}
\affiliation{%
  \institution{CISPA Helmholtz Center for Information Security}
  \city{Saarbr\"{u}cken}
  \country{Germany}}
\email{vladimir.krsmanovic@cispa.de}

\author{Tobias Ladner}
\affiliation{%
  \institution{Technical University of Munich}
  \city{Munich}
  \country{Germany}}
\email{tobias.ladner@tum.de}

\author{Bernd Finkbeiner}
\affiliation{%
  \institution{CISPA Helmholtz Center for Information Security}
  \city{Saarbr\"{u}cken}
  \country{Germany}}
  \affiliation{%
  \institution{Technical University of Munich}
  \city{Munich}
  \country{Germany}}
\email{finkbeiner@cispa.de}

\renewcommand{\shortauthors}{Krsmanovic, Ladner, and Finkbeiner}
\begin{abstract}
Reachability analysis is a popular technique for verifying the safety of a dynamical system under uncertainty, where the uncertainty is often represented by zonotopes. 
As the size of the zonotope representation typically grows as the reachability analysis progresses, it must be repeatedly reduced at the cost of additional conservatism. 
While various methods exist to reduce the size of the representation, combined advantages are not fully leveraged as only a single reduction method is typically used throughout the analysis. 
In this work, we show how dynamically choosing which method to use can yield significantly tighter results, thereby easing verification. 
This is realized by introducing a heuristic, derived from model predictive control, that accounts for upcoming propagation of the current zonotope.
We argue that order reduction is not merely a post-processing step after propagation but also a pre-processing step that shapes how the zonotope evolves under the system dynamics.
We evaluate our approach, \method{}, on nonlinear benchmarks taken from the ARCH-COMP,
where we demonstrate that \method{} produces tighter zonotopes at a negligible performance overhead.
\end{abstract}





\maketitle


\section{Introduction}

A zonotope~\cite{DBLP:conf/hybrid/Girard05} is a convex, centrally symmetric, continuous set that can be represented with only a few parameters even in high-dimensional spaces.
They are a crucial element in reachability analysis of dynamical systems~\cite{DBLP:conf/hybrid/Girard05,althoff2010reachability}, neural network verification~\cite{singh2018fast}, and runtime monitoring under uncertainty~\cite{DBLP:journals/corr/cuttingcorners, DBLP:conf/rv/FinkbeinerFKK24}, as the required operations are efficient to compute for zonotopes.

In these applications, the representation of a zonotope usually grows in size, i.e., the number of generators that span the zonotope grows.
To remain computationally efficient, zonotope reduction methods try to find another zonotope that encloses the original zonotope with fewer generators, where a more aggressive reduction usually leads to greater conservatism in the overarching goal of verifying a safety-critical system.

\begin{figure}[t]
    \centering
    \input{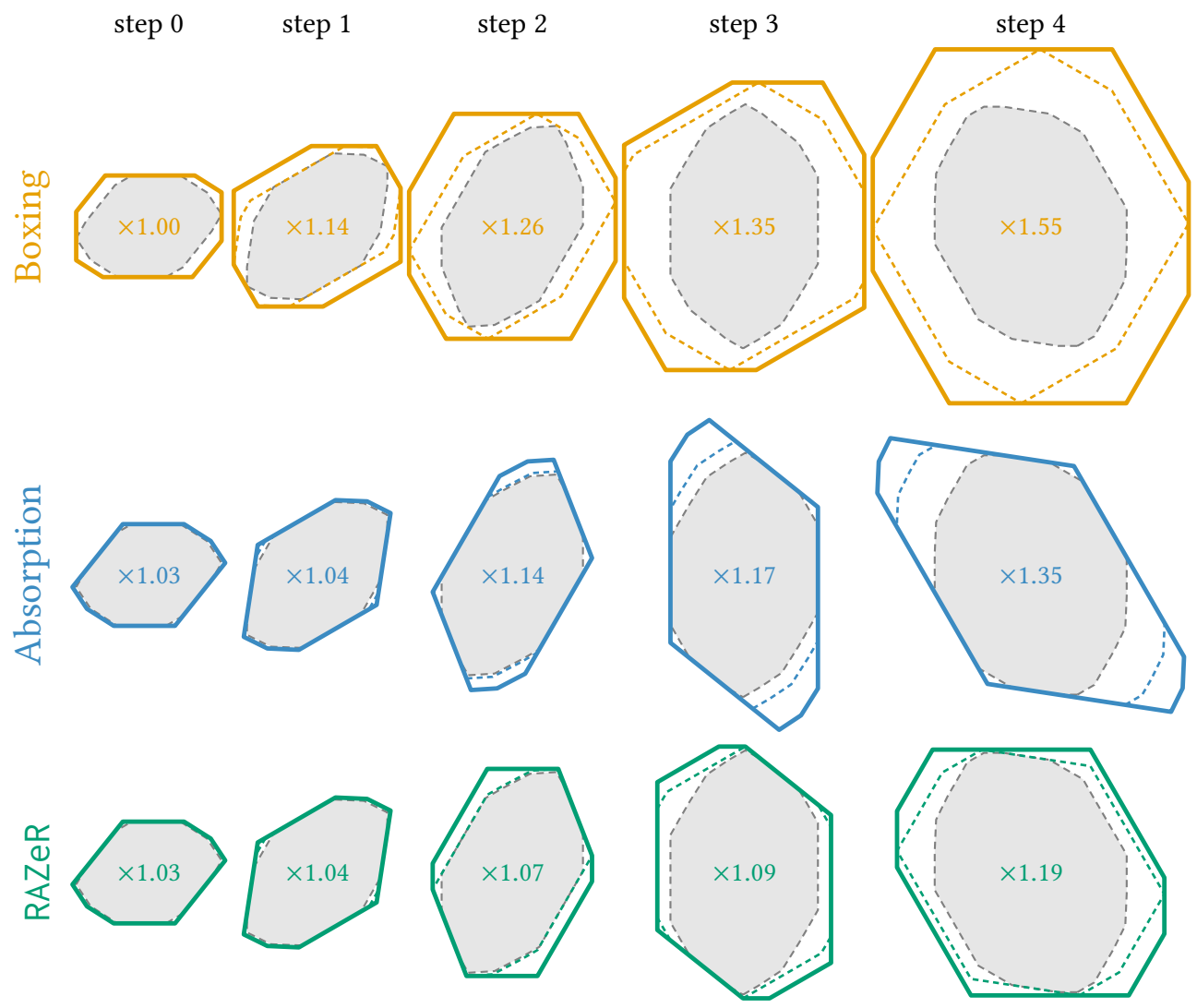}
    \caption{Illustration of how RAZeR improves zonotope reductions in sequential applications by balancing between absorption and boxing of surplus generators in each step.}
    \label{fig:razer}
\end{figure}

\paragraph{Related work.}
Zonotope reduction methods have long been a subject of study~\cite{kopetzki2017methods}.
Common methods simplify the representation of a zonotope by enclosing some generators with an axis-aligned box~\cite{DBLP:conf/hybrid/Girard05,DBLP:conf/eucc/Combastel03,kopetzki2017methods},
or absorb them into the remaining generators~\cite{DBLP:journals/automatica/scott}~ (\cref{fig:razer}).
More advanced methods use generator clustering~\cite{kopetzki2017methods} and constrained optimization~\cite{sadraddini2019linear}.

These reduction methods are static and focus on computing an enclosure independent of the application.
In such applications, a reduction method should not only tightly enclose a given zonotope but also account for its evolution in subsequent steps. \pagebreak

This led to algorithms automatically tuning hyperparameters to balance conservatism across current and future time steps~\cite{wetzlinger2020adaptive}; however, the choice of reduction method was not considered.
This choice can be viewed as an optimal control problem over a receding horizon~\cite{krsmanovic2026predictive}.
While this has been shown to be effective, it comes with a higher computational cost and greater system complexity. 

\paragraph{Contribution.} In this work, we continue this line of research. In summary, our main contributions are:
\begin{itemize}
    \item We derive a computationally cheap heuristic resembling the future cost of different reduction methods in sequential applications (\cref{sec:costofred}).
    \item We use this heuristic to develop our meta-reduction method RAZeR, which dynamically chooses promising reduction methods on the generator level (\cref{sec:razer}).
    \item We evaluate RAZeR on nonlinear benchmarks from ARCHCOMP, resulting in substantially tighter zonotopes at minor runtime overhead (\cref{sec:eval}).
\end{itemize}

\pagebreak
\section{Background}
\label{sec:background}

\paragraph{Notation.}
We denote vectors with lowercase letters, matrices with uppercase letters, and sets with calligraphic letters.
For $n\in\N$, we denote the $n$-dimensional identity matrix by $I_n$ and use $[n]=\{1,\ldots,n\}$.
Given a vector $v\in\R^n$, the operation $\diag{v}$ creates a matrix with the entries of $v$ on its main diagonal.
The concatenation of two matrices $M_1\in\R^{n\times m_1}$, $M_2\in\R^{n\times m_2}$ is denoted by $[M_1 M_2]\in\R^{n\times (m_1+m_2)}$. 
An interval with bounds $l,u\in\R^n$ is denoted by $[l,u]$, where $l\leq u$ holds element-wise.
The bold letter $\mathbf{0}$ denotes a vector with all zeros of proper dimensions.

\begin{figure}[t]
\centering
\definecolor{girard}{HTML}{E69F00}
\definecolor{scott}{HTML}{3B8BC2}
\definecolor{razer}{HTML}{009E73}
\definecolor{combastel}{HTML}{D4B300}
\definecolor{pca}{HTML}{D55E00}
\definecolor{exactfill}{HTML}{E6E6E6}
\definecolor{exactedge}{HTML}{808080}

\begin{tikzpicture}[font=\scriptsize, exact/.style={fill=exactfill, draw=exactedge, line width=.4pt, dash pattern=on 1.5pt off 1pt},
  pre/.style={line width=.5pt, dash pattern=on 1.2pt off .9pt},
  red/.style={line width=.9pt, line join=round}]
\begin{scope}[shift={(1.4,0)}, x=0.255cm, y=0.255cm]
\draw[fill=exactfill, draw=exactedge, line width=.4pt] (-1.5,-3.1) -- (2.5,-3.1) -- (4.9,-0.1) -- (3.7,1.7) -- (1.5,3.1) -- (-2.5,3.1) -- (-4.9,0.1) -- (-3.7,-1.7) -- cycle;
\draw[-stealth, line width=.7pt] (0,0) -- (2,0);
\node at (2.5,0) {$g_2^*$};
\draw[-stealth, line width=.7pt] (0,0) -- (1.2,1.5);
\node at (1.35,1.95) {$g_1^*$};
\draw[-stealth, line width=.7pt] (0,0) -- (1.1,-0.7);
\node at (1.4,-1.1) {$g_3$};
\draw[-stealth, line width=.7pt] (0,0) -- (-0.6,0.9);
\node at (-1.05,1.05) {$g_4$};
\fill (0,0) circle (.9pt);
\end{scope}
\node[anchor=base] at (1.4,0.909) {\small $\mathcal{Z}$};
\begin{scope}[shift={(4.2,0)}, x=0.182cm, y=0.182cm]
\draw[fill=exactfill, draw=exactedge, line width=.4pt] (-1.5,-3.1) -- (2.5,-3.1) -- (4.9,-0.1) -- (3.7,1.7) -- (1.5,3.1) -- (-2.5,3.1) -- (-4.9,0.1) -- (-3.7,-1.7) -- cycle;
\draw[exactedge, line width=.4pt, dash pattern=on 1.5pt off 1pt] (-4.9,-3.1) rectangle (4.9,3.1);
\draw[red, girard] (-4.9,-3.1) -- (4.9,-3.1) -- (4.9,3.1) -- (-4.9,3.1) -- cycle;
\end{scope}
\node[anchor=base, text=girard] at (4.2,0.909) {\small Boxing (\girard{})};
\node[anchor=north, text=girard] at (4.2,-0.613) {hull unchanged};
\begin{scope}[shift={(7,0)}, x=0.182cm, y=0.182cm]
\draw[fill=exactfill, draw=exactedge, line width=.4pt] (-1.5,-3.1) -- (2.5,-3.1) -- (4.9,-0.1) -- (3.7,1.7) -- (1.5,3.1) -- (-2.5,3.1) -- (-4.9,0.1) -- (-3.7,-1.7) -- cycle;
\draw[exactedge, line width=.4pt, dash pattern=on 1.5pt off 1pt] (-4.9,-3.1) rectangle (4.9,3.1);
\draw[red, scott] (-7.46,-3.1) -- (2.5,-3.1) -- (7.46,3.1) -- (-2.5,3.1) -- cycle;
\end{scope}
\node[anchor=base, text=scott] at (7,0.909) {\small Absorption (\scott{})};
\node[anchor=north, text=scott] at (7,-0.613) {total radius $+2.6$};
\end{tikzpicture}
\caption{Comparison of boxing and absorption methods}
\label{fig:reduce}
\end{figure}
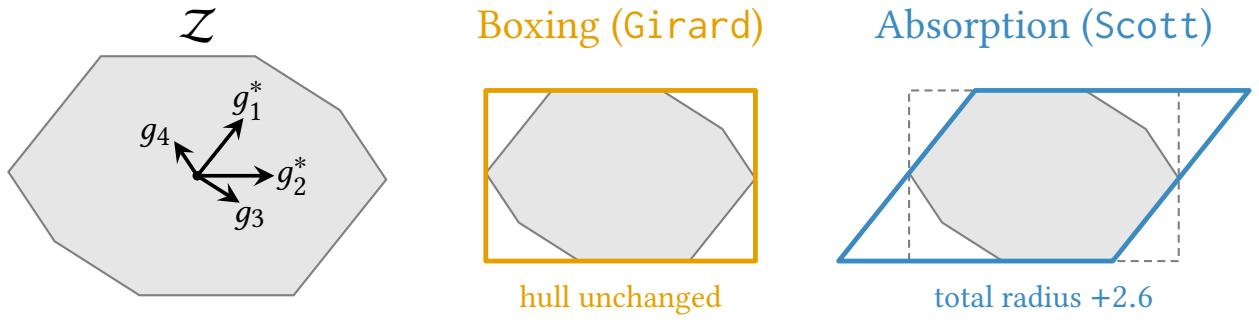

\paragraph{Set-based computing.}
Zonotopes are a set representation commonly used in reachability analysis due to their efficient computation of the required operations.
\begin{definition}[Zonotope~\cite{DBLP:conf/hybrid/Girard05}]
 \label[definition]{def:zonotope}
Given a center $c\in\R^n$ and a generator matrix $G\in\R^{n\times h}$, a zonotope is defined by
\[
  \Z=\shortZ{c}{G}=\defZ.
\]
\end{definition}
The most common operations in reachability analysis are the linear map and the Minkowski sum.
For a matrix $M\in\R^{m\times n}$, and zonotopes $\Z_1\shortZ{c_1}{G_1}$, $\Z_2\shortZ{c_2}{G_2}\subset\R^n$, these are computed by
\begin{align}
\label{eq:ops}
\begin{split}
  M\Z_1&= \left\{ Mz_1 \ |\ z_1\in\Z_1 \right\}=\shortZ{Mc_1}{MG_1},\\
  \Z_1\oplus\Z_2&=\left\{ z_1{+}z_2 \ |\ z_1\in\Z_1,\,z_2\in\Z_2 \right\}=\shortZ{c_1{+}c_2}{[G_1\ G_2]}.
  \end{split}
\end{align}
The number of generators $h$ typically grows during reachability analysis, where we use the order of a zonotope $\Zorder=h/n\geq 0$ as a measure with respect to the system dimension $n$. 
To maintain computational efficiency, we have to repeatedly apply an order reduction~\cite{kopetzki2017methods} at the cost of additional outer approximation. 
For a given zonotope $\Z=\shortZ{c}{G}\subset\R^n$,
\begin{equation} 
\label{eq:reduce}
    \widehat{\Z}=\reduce{\Z}{method}{\Zorder}\supseteq\Z
\end{equation}
ensures that $\widehat{\Z}$ has at most $n\Zorder$ generators.
There exist various order reduction methods~\cite{kopetzki2017methods}, which we broadly categorize into boxing and absorption methods (\cref{fig:reduce}):

\girard{}~\cite{DBLP:conf/hybrid/Girard05} splits the generator matrix $G$ into kept generators $G_K\in\R^{n\times (\Zorder-1)n}$ and surplus generators $G_S\in\R^{n\times (h-(\Zorder-1)n)}$, and computes a box enclosure of $G_S$, resulting in at most $n$ new generators:
\begin{align}
    \begin{split}
            \widehat{\Z}&=\reduce{\Z}{\girard}{\Zorder}= \shortZ{c}{G_K} \oplus \interval{\shortZ{\mathbf{0}}{G_S}} \supseteq \Z,
    \end{split}
\end{align}
where 
\begin{align}
\label{eq:rad}
\begin{split}
  \interval{\Z}&=\shortZ{c}{\diag{\rad(\Z)}} \\
  \text{with } \rad(\Z)&=\sum_{i=1}^h\abs{g_i}\in\R^n.
  \end{split}
\end{align}
Similarly, \combastel{}~\cite{DBLP:conf/eucc/Combastel03} boxes the generators of the smallest Euclidean norm, and \pca{}~\cite{kopetzki2017methods} boxes the surplus in the principal frame of the surplus generators instead of the coordinate frame.

In contrast, \scott{}~\cite{DBLP:journals/automatica/scott} absorbs surplus generators by scaling the kept generators accordingly. 
It uses Gauss--Jordan elimination to determine a subset of generators $G^*=[g_1^* \cdots g_n^*]\in\R^{n\times n}$ spanning an invertible basis; the remaining generators are again partitioned into some generators $G'_K$ that are kept unchanged and the remaining $G'_S$ are subject to the reduction. 
Without loss of generality, let
\begin{equation}
    \label{eq:scott-decomposition}
    G=[G^*\ G'_K\ G'_S] = [g^*_1 \cdots g_n^*\ \ g_{n+1} \cdots g_{\Zorder n}\ \ g_{\Zorder n+1}  \cdots g_h].    
\end{equation}
As $G^*$ forms a basis, we can express $G'_S$ in terms of $G^*$:
\begin{equation}
  V_S=(G^*)^{-1} G'_S,
  \label{eq:scott-transform}
\end{equation}
which gets absorbed into the basis by computing:
\begin{equation}
  \widehat{G}^* \gets G^*\left(I_n + \diag{\rad(\shortZ{\mathbf{0}}{V_S})}\right).
  \label{eq:scott-absorb}
\end{equation}
This effectively computes a box enclosure in the coordinate frame spanned by $G^*$.
As this box is aligned with $G^*$, each generator of the box can be summed with the respective generator of $G^*$.
Thus, it holds:
\begin{equation}
    \widehat{\Z}=\reduce{\Z}{\scott}{\Zorder} = \shortZ{c}{\left[\widehat{G}^*\ G'_K\right]} \supseteq \Z.
\end{equation}

\paragraph{Reachability analysis}

We present an intentionally simplified algorithm to conduct reachability analysis of a nonlinear discrete-time system in \cref{alg:reach}.
This high-level description is sufficient for this work; we defer interested readers to \cite{althoff2010reachability} for more information.
The initial set $\contSet{X}_0\in\R^n$ is iteratively propagated forward in time, where the nonlinear dynamic $x_{k+1}=f(x_k)$ is linearized at each time point $k$, yielding a linear propagation map $M\in\R^{n\times n}$.
This can also be generalized to continuous time systems, i.e., 
let $A\in\R^{n\times n}$ be the matrix describing the linearized dynamics, then $M=e^{A_k\Delta t}$.
To account for the linearization error, the set is expanded after each propagation using \eqref{eq:ops},
where the Minkowski sum results in additional generators.
Thus, we are required to repeatedly reduce the order of $\contSet{X}_k$ to remain computationally feasible.

\begin{algorithm}[h]
    \caption{Simplified reachability analysis}
    \label{alg:reach}
    \DontPrintSemicolon\SetAlgoNoEnd\SetArgSty{textnormal}
    \SetKwComment{tcp}{$\triangleright$ }{}\SetCommentSty{textit}
    \KwIn{System dynamics $f(x)$; initial set $\contSet{X}_0$; max. order $\Zorder$; time horizon $k_{\max}$}
    \KwOut{Reachable set $\contSet{R}$}
      \For{$k \in [k_{\max}]$}{
            Determine linearization point $x_k^*$\;
            Compute linear propagation map $M$ using $x_k^*$\; \label{alg-line:compute-map}
            Compute linearization error $\contSet{E}$\;
            $\contSet{X}_k\gets M\contSet{X}_{k-1} \oplus \contSet{E}$ \tcp*{\eqref{eq:ops}} \label{alg-line:propagation} 
            $\contSet{X}_k \gets \reduce{\contSet{X}_k}{method}{\Zorder}$\tcp*{\eqref{eq:reduce}} \label{alg-line:reduce}
      }
    \Return $\contSet{R} \gets \bigcup_k\contSet{X}_k$\; 
\end{algorithm}

\section{Cost of Reduction over Time}
\label{sec:costofred}

We measure the cost of the reduction via the volume of the resulting zonotope. 
As the volume of zonotopes is expensive to compute in higher dimensions~\cite{dyer1998complexity},
we approximate it with the volume of the interval hull as computed in \eqref{eq:rad}.
This slightly benefits reduction methods based on boxing;
however, as we are considering the entire reachability analysis over time, 
we assume this effect to be negligible if another method clearly outperforms boxing methods.

\begin{figure}[t]
\centering
\definecolor{girard}{HTML}{E69F00}
\definecolor{scott}{HTML}{3B8BC2}
\definecolor{razer}{HTML}{009E73}
\definecolor{combastel}{HTML}{D4B300}
\definecolor{pca}{HTML}{D55E00}
\definecolor{exactfill}{HTML}{E6E6E6}
\definecolor{exactedge}{HTML}{808080}

\begin{tikzpicture}[font=\scriptsize, exact/.style={fill=exactfill, draw=exactedge, line width=.4pt, dash pattern=on 1.5pt off 1pt},
  pre/.style={line width=.5pt, dash pattern=on 1.2pt off .9pt},
  red/.style={line width=.9pt, line join=round}]
\draw[exact] (1.03,-0.354) -- (1.487,-0.354) -- (1.761,-0.011) -- (1.624,0.194) -- (1.373,0.354) -- (0.916,0.354) -- (0.642,0.011) -- (0.779,-0.194) -- cycle;
\draw[red, girard] (0.642,-0.354) -- (1.761,-0.354) -- (1.761,0.354) -- (0.642,0.354) -- cycle;
\node[anchor=north, text=girard] at (1.201,-1.245) {$\times1.00$};
\draw[exact] (2.914,-0.392) -- (2.982,-0.392) -- (3.378,-0.164) -- (3.444,0.27) -- (3.222,0.379) -- (2.925,0.392) -- (2.857,0.392) -- (2.461,0.164) -- (2.395,-0.27) -- (2.617,-0.379) -- cycle;
\draw[pre, girard] (2.578,-0.586) -- (2.646,-0.586) -- (3.615,-0.027) -- (3.261,0.586) -- (3.193,0.586) -- (2.224,0.027) -- cycle;
\draw[red, girard] (2.224,-0.586) -- (3.615,-0.586) -- (3.615,0.586) -- (2.224,0.586) -- cycle;
\node[anchor=north, text=girard] at (2.919,-1.245) {$\times1.40$};
\draw[exact] (4.362,-0.496) -- (4.431,-0.496) -- (4.677,-0.48) -- (4.941,-0.342) -- (5,-0.308) -- (5.229,0.087) -- (5.069,0.496) -- (5.001,0.496) -- (4.754,0.48) -- (4.49,0.342) -- (4.431,0.308) -- (4.203,-0.087) -- cycle;
\draw[pre, girard] (4.372,-0.855) -- (4.441,-0.855) -- (5.645,-0.16) -- (5.059,0.855) -- (4.991,0.855) -- (3.786,0.16) -- cycle;
\draw[red, girard] (3.786,-0.855) -- (5.645,-0.855) -- (5.645,0.855) -- (3.786,0.855) -- cycle;
\node[anchor=north, text=girard] at (4.716,-1.245) {$\times1.77$};
\draw[exact] (6.991,-0.606) -- (7.059,-0.606) -- (7.119,-0.572) -- (7.324,-0.435) -- (7.484,-0.184) -- (7.518,-0.124) -- (7.518,0.332) -- (7.176,0.606) -- (7.107,0.606) -- (7.048,0.572) -- (6.842,0.435) -- (6.683,0.184) -- (6.648,0.124) -- (6.648,-0.332) -- cycle;
\draw[pre, girard] (6.672,-1.205) -- (6.74,-1.205) -- (8.35,-0.276) -- (7.495,1.205) -- (7.426,1.205) -- (5.816,0.276) -- cycle;
\draw[red, girard] (5.816,-1.205) -- (8.35,-1.205) -- (8.35,1.205) -- (5.816,1.205) -- cycle;
\node[anchor=north, text=girard] at (7.083,-1.245) {$\times2.38$};
\node[rotate=90, anchor=south, text=girard] at (0.3,0) {\small Boxing};
\draw[exact] (1.03,-3.022) -- (1.487,-3.022) -- (1.761,-2.68) -- (1.624,-2.474) -- (1.373,-2.315) -- (0.916,-2.315) -- (0.642,-2.657) -- (0.779,-2.862) -- cycle;
\draw[red, scott] (0.35,-3.022) -- (1.487,-3.022) -- (2.053,-2.315) -- (0.916,-2.315) -- cycle;
\node[anchor=north, text=scott] at (1.201,-3.751) {$\times1.32$};
\draw[exact] (2.914,-3.06) -- (2.982,-3.06) -- (3.378,-2.832) -- (3.444,-2.399) -- (3.222,-2.289) -- (2.925,-2.276) -- (2.857,-2.276) -- (2.461,-2.505) -- (2.395,-2.938) -- (2.617,-3.048) -- cycle;
\draw[pre, scott] (2.325,-3.4) -- (2.393,-3.4) -- (3.378,-2.832) -- (3.514,-1.936) -- (3.446,-1.936) -- (2.461,-2.505) -- cycle;
\draw[red, scott] (2.318,-3.444) -- (3.378,-2.832) -- (3.521,-1.893) -- (2.461,-2.505) -- cycle;
\node[anchor=north, text=scott] at (2.919,-3.751) {$\times1.50$};
\draw[exact] (4.362,-3.164) -- (4.431,-3.164) -- (4.677,-3.148) -- (4.941,-3.011) -- (5,-2.976) -- (5.229,-2.581) -- (5.069,-2.173) -- (5.001,-2.173) -- (4.754,-2.189) -- (4.49,-2.326) -- (4.431,-2.36) -- (4.203,-2.756) -- cycle;
\draw[pre, scott] (4.548,-3.641) -- (4.617,-3.641) -- (5.229,-2.581) -- (4.883,-1.696) -- (4.814,-1.696) -- (4.203,-2.756) -- cycle;
\draw[red, scott] (4.576,-3.711) -- (5.229,-2.581) -- (4.855,-1.625) -- (4.203,-2.756) -- cycle;
\node[anchor=north, text=scott] at (4.716,-3.751) {$\times1.54$};
\draw[exact] (6.991,-3.274) -- (7.059,-3.274) -- (7.119,-3.24) -- (7.324,-3.103) -- (7.484,-2.852) -- (7.518,-2.793) -- (7.518,-2.336) -- (7.176,-2.062) -- (7.107,-2.062) -- (7.048,-2.097) -- (6.842,-2.234) -- (6.683,-2.485) -- (6.648,-2.544) -- (6.648,-3) -- cycle;
\draw[pre, scott] (7.449,-3.641) -- (7.518,-3.641) -- (7.518,-2.336) -- (6.717,-1.695) -- (6.648,-1.695) -- (6.648,-3) -- cycle;
\draw[red, scott] (7.518,-3.696) -- (7.518,-2.336) -- (6.648,-1.641) -- (6.648,-3) -- cycle;
\node[anchor=north, text=scott] at (7.083,-3.751) {$\times1.41$};
\node[rotate=90, anchor=south, text=scott] at (0.3,-2.668) {\small Absorption};
\node[anchor=base] (s0) at (1.201,1.325) {step 0};
\node[anchor=base] (s1) at (2.919,1.325) {step 1};
\node[anchor=base] (s2) at (4.716,1.325) {step 2};
\node[anchor=base] (s3) at (7.083,1.325) {step 3};
\draw[-stealth, line width=.4pt, shorten <=3pt, shorten >=3pt] (s0.mid east) -- node[above=-1pt, font=\scriptsize] {$M$} (s1.mid west);
\draw[-stealth, line width=.4pt, shorten <=3pt, shorten >=3pt] (s1.mid east) -- node[above=-1pt, font=\scriptsize] {$M$} (s2.mid west);
\draw[-stealth, line width=.4pt, shorten <=3pt, shorten >=3pt] (s2.mid east) -- node[above=-1pt, font=\scriptsize] {$M$} (s3.mid west);
\end{tikzpicture}
\caption{Wrapping effect with an aggressive order reduction to $\Zorder=1$ after each step.}
\label{fig:wrap}
\end{figure}
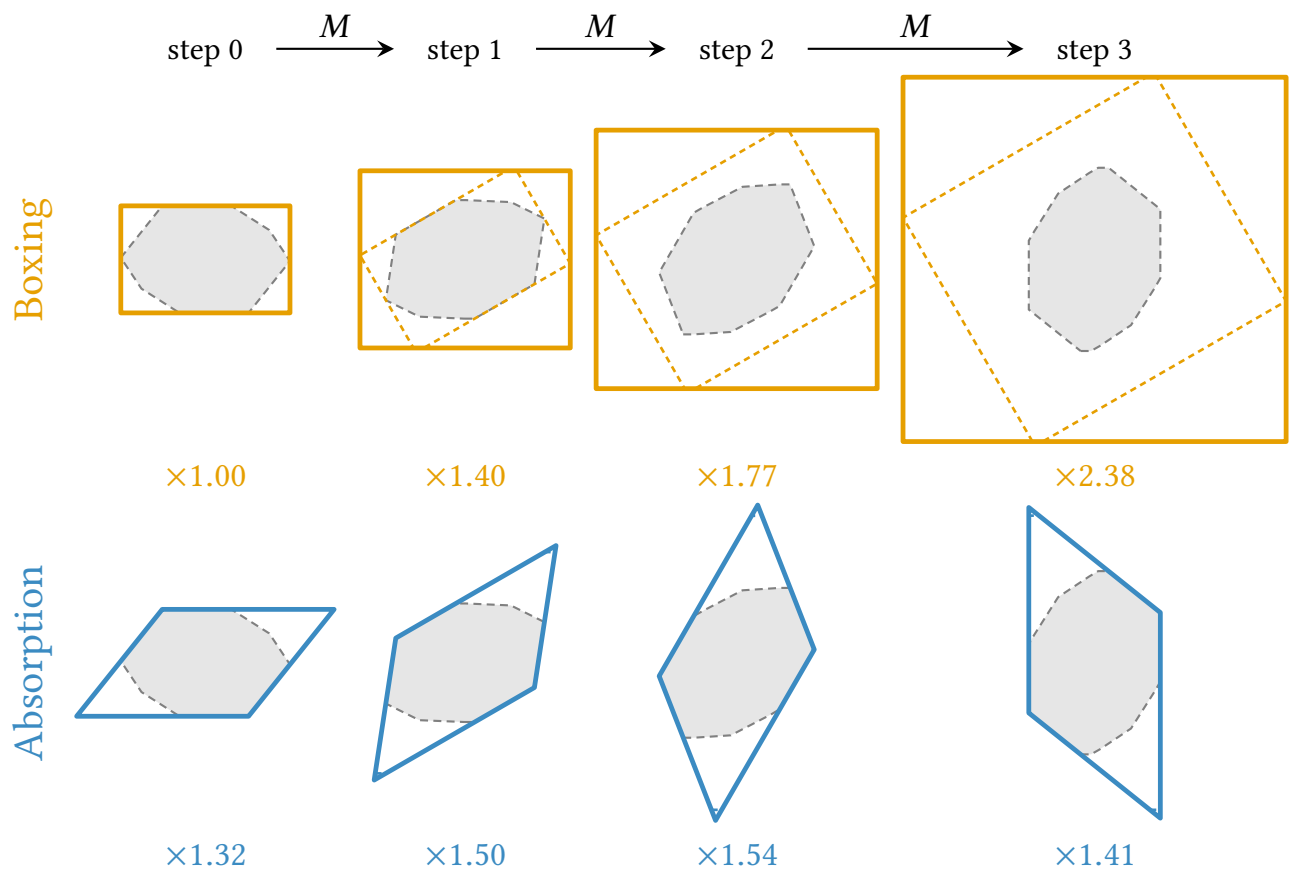

In particular, boxing methods are subject to the wrapping effect if applied repeatedly, as illustrated in \cref{fig:wrap}.
Thus, it is reasonable to measure the cost of reduction over time as well.
We compute this cost greedily by only looking one time step into the future,
effectively bounding the difference of applying \eqref{eq:rad} before and after this step.
\begin{definition}[Wrapping cost]
 \label[definition]{def:defect}
For a linear map $M\in\R^{m\times n}$ and a generator $g\in\R^n$, 
the wrapping cost is given by:
\[
  \wrap(M,g)=\norm{\abs{M}\abs{g}}_1-\norm{\abs{Mg}}_1 \ge 0.
\]

\end{definition}  
The overall cost for applying a boxing method is thus given by:
\begin{definition}[Boxing cost] \label[definition]{def:boxcost}
For a linear map $M\in\R^{m\times n}$ and a generator $g\in\R^n$, 
the cost of boxing $g$ is given by:
\begin{equation*}
  \boxcost_M(g)=\wrap(M,g)=\norm{\abs{M}\abs{g}}_1-\norm{Mg}_1 \ge 0.
  \label{eq:box-cost}
\end{equation*}
\end{definition}   
The cost for absorption methods works analogously; we just have to respect the basis $G^*$ while doing so:
\begin{definition}[Absorption cost]
 \label[definition]{def:abscost}
For a linear map $M\in\R^{m\times n}$, a basis $G^*\in\R^{n\times n}$, and a generator $g\in\R^n$ with $v=(G^*)^{-1}g$, the absorption cost of $g$ is given by:
\begin{equation*}
  \abscost_M(g)=\wrap(MG^*,v)=\norm{\abs{MG^*}\abs{v}}_1-\norm{Mg}_1 \ge 0.
  \label{eq:abs-cost}
\end{equation*}
\end{definition}

Using these costs, we provide further insights into how a model predictive control controller~\cite{krsmanovic2026predictive} chooses which order-reduction method to use.
In particular, we notice that \scott{} is preferred as long as the cumulative absorption cost over all generators is below some threshold,
and falls back to boxing methods above that threshold.
More formally, let the cumulative absorption cost be:
\begin{equation}
  \abscostcum=\frac{\sum_k\abscost_{I_n}(g_k)}{\sum_k\norm{g_k}_1},
  \label{eq:etabar}
\end{equation}
where we used the identity $I_n$ in the numerator, since model predictive control does not directly account for map propagation to future time steps. 

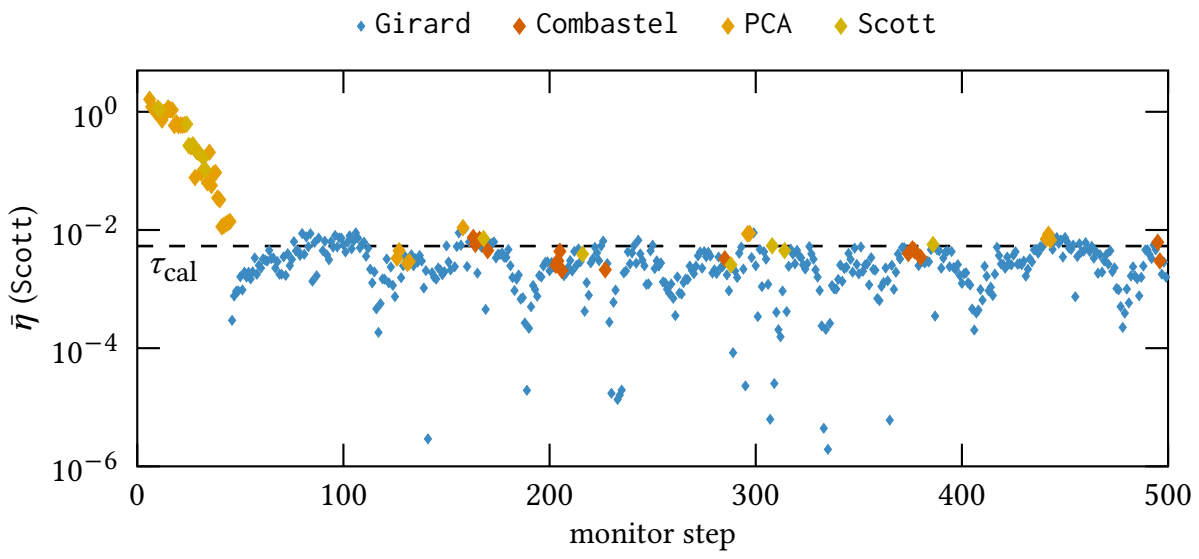
\begin{figure}[h]
\centering
\definecolor{girard}{HTML}{E69F00}
\definecolor{scott}{HTML}{3B8BC2}
\definecolor{razer}{HTML}{009E73}
\definecolor{combastel}{HTML}{D4B300}
\definecolor{pca}{HTML}{D55E00}
\definecolor{exactfill}{HTML}{E6E6E6}
\definecolor{exactedge}{HTML}{808080}

\pgfplotsset{compat=1.18}
\begin{tikzpicture}[font=\scriptsize]
\begin{axis}[width=8.4cm, height=4.2cm, scale only axis=false, ymode=log, xmin=0, xmax=500, ymin=1e-6, ymax=5,
  axis lines=box, axis line style={-, line width=.4pt}, tick style={line width=.4pt, black}, tick align=inside,
  ytick={1e-6,1e-4,1e-2,1}, minor ytick={}, xtick={0,100,...,500}, tick label style={font=\scriptsize},
  xlabel={monitor step}, ylabel={$\bar\eta$ (\scott{})}, label style={font=\scriptsize},
  xlabel style={yshift=3pt}, ylabel style={yshift=-4pt},
  legend columns=4, legend style={draw=none, fill=none, font=\scriptsize, at={(0.5,1.02)}, anchor=south,
    /tikz/every even column/.append style={column sep=6pt}}, legend cell align=left,
  clip=false]
\addplot[only marks, mark=diamond*, mark size=0.9pt, color=scott, mark options={line width=.2pt}] coordinates {
  (46,0.0002957) (47,0.0007637) (48,0.0009189) (49,0.000931) (50,0.001621) (51,0.0009666) (52,0.001933) (53,0.001594)
  (54,0.001922) (55,0.001486) (56,0.001962) (57,0.003365) (58,0.001778) (59,0.002524) (60,0.00227) (61,0.003229)
  (62,0.003099) (63,0.002983) (64,0.004515) (65,0.003277) (66,0.003357) (67,0.002269) (68,0.002913) (69,0.001777)
  (70,0.001785) (71,0.003519) (72,0.001702) (73,0.00398) (74,0.002493) (75,0.003182) (76,0.006357) (77,0.004408)
  (78,0.004449) (79,0.003724) (80,0.008709) (81,0.003879) (82,0.007636) (83,0.0052) (84,0.008672) (85,0.00136)
  (86,0.006159) (87,0.001684) (88,0.007281) (89,0) (90,0.004462) (91,0.006442) (92,0.00379) (93,0.00314)
  (94,0.006359) (95,0.005141) (96,0.008065) (97,0.00517) (98,0.007366) (99,0.007307) (100,0.005619) (101,0.004251)
  (102,0.006228) (103,0.007958) (104,0.004798) (105,0.006526) (106,0.008952) (107,0.007348) (108,0.004994) (109,0.005764)
  (110,0.006216) (111,0.003596) (112,0.004921) (113,0.002168) (114,0.001272) (115,0.001952) (116,0.0004629) (117,0.0001849)
  (118,0.0005666) (119,0.001485) (120,0.001464) (121,0.003227) (122,0.001277) (123,0.002368) (124,0.001156) (125,0.0007464)
  (128,0.001558) (129,0.002488) (130,0.003535) (133,0.003812) (134,0.002488) (135,0.003255) (136,0.002856) (137,0.00371)
  (138,0.002983) (139,0.001042) (140,0.002053) (141,2.916e-06) (142,0) (143,0.001891) (144,0.001576) (145,0.001478)
  (146,0.002273) (147,0.00211) (148,0.001758) (149,0.001965) (150,0.005387) (151,0.002869) (152,0.002104) (153,0)
  (154,0.002391) (155,0.005252) (156,0.008945) (157,0.005526) (159,0.01061) (160,0.002409) (161,0.006268) (162,0)
  (165,0.001861) (167,0.005092) (169,0.0004528) (171,0.005867) (172,0.005932) (173,0.006286) (174,0.003371) (175,0.003894)
  (176,0.003856) (177,0.004061) (178,0.003447) (179,0.003438) (180,0.00467) (181,0.002483) (182,0.001569) (183,0.0009447)
  (184,0.001477) (185,0.001732) (186,0.0006756) (187,0.0007165) (188,0.0002668) (189,1.934e-05) (190,0.0002166) (191,0.0005023)
  (192,0.00115) (193,0.0009827) (194,0.0007568) (195,0.001693) (196,0.001605) (197,0.002471) (198,0.002849) (199,0.002882)
  (200,0.002039) (201,0.002815) (202,0.002384) (207,0.001685) (208,0.001851) (209,0.002333) (210,0.00304) (211,0.002025)
  (212,0.004337) (213,0.002509) (214,0.004775) (215,0.002915) (217,0.000421) (218,0.001285) (219,0.0007862) (220,0.002788)
  (221,0.003256) (222,0.005074) (223,0.004096) (224,0.003018) (225,0.005365) (226,0.006488) (228,0.002702) (229,0.0002756)
  (230,1.723e-05) (231,0.0005973) (232,0.0009617) (233,1.356e-05) (234,1.601e-05) (235,1.953e-05) (236,0.001883) (237,0.002023)
  (238,0.004283) (239,0.002675) (240,0.004099) (241,0.005462) (242,0.002738) (243,0.006707) (244,0.003932) (245,0.003081)
  (246,0.002474) (247,0.001196) (248,0.001225) (249,0.00444) (250,0.00553) (251,0.002225) (252,0.003005) (253,0.002451)
  (254,0.002765) (255,0.001317) (256,0.001282) (257,0.0009911) (258,0.0008547) (259,0.001219) (260,0.0006969) (261,0.0003562)
  (262,0.001604) (263,0.000858) (264,0.0008221) (265,0.001292) (266,0.001478) (267,0.001454) (268,0.002452) (269,0.002076)
  (270,0.001826) (271,0.002441) (272,0.003653) (273,0.004224) (274,0.002256) (275,0.004307) (276,0.003526) (277,0.004185)
  (278,0.003628) (279,0.001871) (280,0.002116) (281,0.001835) (282,0.003484) (283,0.002293) (284,0.00241) (286,0.00174)
  (287,0.0009373) (289,8.37e-05) (290,0.002024) (291,0.001569) (292,0.004032) (293,0.005147) (294,0.002177) (295,2.297e-05)
  (298,0.003416) (299,0.008957) (300,0.002485) (301,0.0003427) (302,0.00468) (303,0.001102) (304,0.004287) (305,0.003208)
  (306,0.001203) (307,6.249e-06) (309,2.521e-05) (310,0.0004045) (311,0.0002057) (312,0.0001557) (313,0.0009277) (315,0.000416)
  (316,0.002426) (317,0.004719) (318,0.003913) (319,0.002679) (320,0.00452) (321,0.003673) (322,0.003381) (323,0.002558)
  (324,0.002182) (325,0.002698) (326,0.005246) (327,0.003937) (328,0.003202) (329,0.002117) (330,0.001595) (331,0.0009465)
  (332,0.0002383) (333,4.403e-06) (334,0.0002079) (335,1.94e-06) (336,0.0002626) (337,0.001145) (338,0.001011) (339,0.001006)
  (340,0.001456) (341,0.002556) (342,0.001355) (343,0.002226) (344,0.00235) (345,0.004014) (346,0.003183) (347,0.001934)
  (348,0.004927) (349,0.002611) (350,0.001812) (351,0.005157) (352,0.002004) (353,0.002552) (354,0.002669) (355,0.001177)
  (356,0.003401) (357,0.002153) (358,0.002088) (359,0.0006937) (360,0.0006393) (361,0.001106) (362,0.001291) (363,0.002278)
  (364,0.001383) (365,6.05e-06) (366,0.002524) (367,0.0009897) (368,0.001929) (369,0.003924) (370,0.00324) (371,0.005028)
  (372,0.005022) (373,0.004066) (375,0.003939) (378,0.003983) (379,0.002496) (381,0.002932) (382,0.002781) (383,0.002981)
  (384,0.002899) (385,0.004654) (387,0.0003492) (388,0.004607) (389,0.00446) (390,0.002255) (391,0.003342) (392,0.00421)
  (393,0.0044) (394,0.002392) (395,0.00431) (396,0.002317) (397,0.002145) (398,0.00259) (399,0.001697) (400,0.001248)
  (401,0.00182) (402,0.001883) (403,0.001273) (404,0.001171) (405,0.0004599) (406,0.0002027) (407,0.0003937) (408,0.0004418)
  (409,0.00098) (410,0.0006508) (411,0.002434) (412,0.001651) (413,0.0009347) (414,0.001078) (415,0.0007795) (416,0.001791)
  (417,0.005103) (418,0.003695) (419,0.002661) (420,0.002617) (421,0.002252) (422,0.003472) (423,0.002684) (424,0.002569)
  (425,0.003754) (426,0.004284) (427,0.002635) (428,0.002722) (429,0.002593) (430,0.002972) (431,0.001659) (432,0.002161)
  (433,0.003239) (434,0.002886) (435,0.00461) (436,0.004662) (437,0.004065) (438,0.005209) (439,0.00513) (440,0.006821)
  (444,0.00469) (445,0.006504) (446,0.007453) (447,0.005941) (448,0.006737) (449,0.003495) (450,0.007463) (451,0.006332)
  (452,0.004943) (453,0.004702) (454,0.005553) (455,0.0007416) (456,0.00356) (457,0.005855) (458,0.004285) (459,0.004362)
  (460,0.006118) (461,0.004879) (462,0.004316) (463,0.005287) (464,0.003342) (465,0.004324) (466,0.001853) (467,0.0038)
  (468,0.001935) (469,0.003002) (470,0.003813) (471,0.002644) (472,0.004456) (473,0.001977) (474,0.002292) (475,0.001024)
  (476,0.0009725) (477,0.0005053) (478,0.0002246) (479,0.0003919) (480,0.001002) (481,0.0005836) (482,0.001664) (483,0.00238)
  (484,0.001658) (485,0.001498) (486,0.0007673) (487,0.001581) (488,0.002455) (489,0.004614) (490,0.004624) (491,0.003867)
  (492,0.003293) (493,0.005548) (494,0.004634) (497,0.001787) (498,0.002865) (499,0.00159)
};
\addlegendentry{\girard{}}
\draw[dashed, line width=.5pt] (axis cs:0,0.005345) -- (axis cs:500,0.005345) node[pos=0, anchor=north west, inner sep=2pt] {$\tau_{\mathrm{cal}}$};
\addlegendentry{\combastel{}}
\addplot[only marks, mark=diamond*, mark size=1.5pt, color=pca, mark options={line width=.2pt}] coordinates {
  (163,0.00748) (164,0.005715) (166,0.006957) (170,0.004491) (203,0.002533) (204,0.002998) (205,0.004352) (206,0.002062)
  (227,0.002119) (285,0.003247) (374,0.003988) (376,0.004795) (377,0.004457) (380,0.003436) (495,0.006189) (496,0.002995)
};
\addlegendentry{\pca{}}
\addplot[only marks, mark=diamond*, mark size=1.5pt, color=girard, mark options={line width=.2pt}] coordinates {
  (6,1.625) (7,1.215) (8,1.144) (9,0.9916) (12,0.726) (13,0.977) (14,0.9761) (15,1.16)
  (16,1.08) (17,1.077) (18,0.5856) (19,0.6515) (20,0.5832) (21,0.5985) (22,0.5908) (23,0.6047)
  (26,0.2569) (28,0.07708) (29,0.2093) (31,0.09228) (34,0.06305) (35,0.2069) (36,0.05674) (37,0.08968)
  (38,0.09346) (39,0.03484) (40,0.03246) (41,0.01136) (42,0.01192) (43,0.01262) (44,0.01327) (45,0.01393)
  (126,0.003333) (127,0.004453) (131,0.002723) (132,0.002911) (158,0.01097) (296,0.008568) (297,0.008749) (441,0.007097)
  (442,0.00845) (443,0.006758)
};
\addlegendentry{\scott{}}
\addplot[only marks, mark=diamond*, mark size=1.5pt, color=combastel, mark options={line width=.2pt}] coordinates {
  (5,0) (10,1.163) (11,1.081) (24,0.6177) (25,0.2681) (27,0.2762) (30,0.2019) (32,0.1663)
  (33,0.1051) (168,0.007155) (216,0.003894) (288,0.002567) (308,0.005398) (314,0.004477) (386,0.005674)
};

\end{axis}
\end{tikzpicture}
\caption{Reduction choices of the Predictive Zonotope Reduction~\cite{krsmanovic2026predictive} are strongly correlated with the absorption cost.}
\label{fig:pzr}
\end{figure}

Using $\abscostcum$, we illustrate this threshold behavior in \cref{fig:pzr} by plotting the actions of the controller in an example reachability run with respect to $\abscostcum$:
Absorption methods work well when the bases are orthogonal to each other, as then the absorption is lossless as it absorbs into a box \eqref{eq:scott-absorb}. However, as monitor updates and reductions can introduce shear into the zonotope, this can, over time, lead to greater error. This slowly increases $\bar\eta$ until the controller detects the increasing cost on the horizon and switches to boxing methods, which then remove the shear and restore the shape, pushing $\bar\eta$ down to levels where absorption again provides more benefit.
Across our reachability runs    , $\abscostcum$ threshold agrees with the chosen method by the controller (\scott{} vs.\ others) with an AUC of 0.85.

\begin{figure*}[b]
\centering
\input{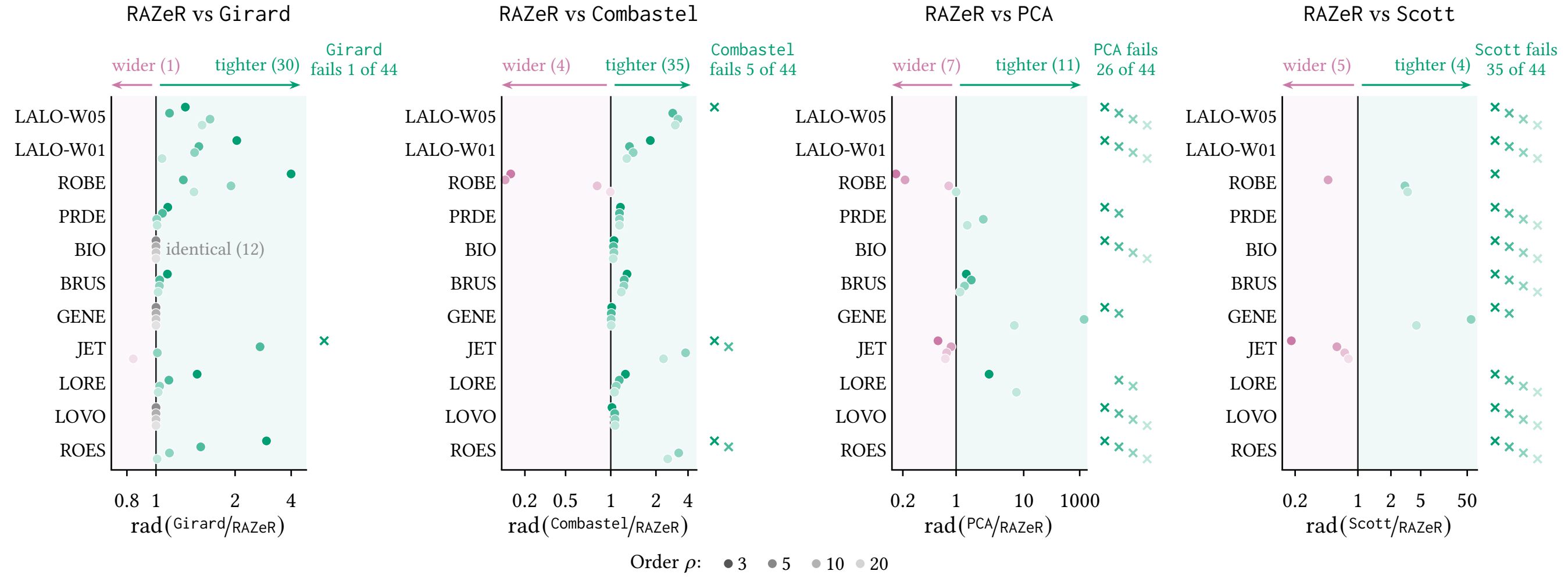}
\caption{
Comparison of \method{} against each established method using~\eqref{eq:eval-metric} with different orders $\Zorder$, on log scale. 
A value above $1$ means that \method{} is tighter than the respective method. 
Runs where the reachability analysis using an established method fails due to a set explosion are shown with a cross ($\times$).
}
\label{fig:dots}
\end{figure*}

\section{Representation Aware Zonotope Reduction}\label{sec:razer}
We build on the observation in \cref{sec:costofred} to develop \method{} 
-- standing for Representation Aware Zonotope Reduction -- 
which balances the costs of absorption and boxing for upcoming propagation steps. It uses this information to intelligently select which surplus generators to absorb and which to box, gaining the benefits of model predictive control while avoiding its overhead.

\begin{algorithm}[h]
    \caption{Order reduction method \method{}}
    \label{alg:razer}
    \DontPrintSemicolon\SetAlgoNoEnd\SetArgSty{textnormal}
    \SetKwComment{tcp}{$\triangleright$ }{}\SetCommentSty{textit}
    \SetKwBlock{AbsorptionPhase}{absorption phase}{}
    \SetKwBlock{BoxingPhase}{boxing phase}{}
    \KwIn{Zonotope $\Z=\shortZ{c}{G}$; max. order $\Zorder$; next map $M$}
    \KwOut{$\widehat{\Z}\supseteq\Z$ with at most $n\Zorder$ generators}
    \lIf{$h\le n\Zorder$}{\Return $\Z$}\label{line:trivial}
    \AbsorptionPhase{
      Determine basis $G^*$ and remainder $G'$ \tcp*{\eqref{eq:scott-decomposition}}
      Compute $b_M(V)$ for $V=(G^*)^{-1}G'$ \tcp*{\cref{def:benefit}, \eqref{eq:scott-transform}}\label{line:benefit}
      Partition $G'$ into $G'_K$ and $G'_S$ using $b_M(V) > 0$ \tcp*{\eqref{eq:scott-decomposition}}
      $\widehat{G}^*\gets$ Absorb $G'_S$ into basis \tcp*{\eqref{eq:scott-absorb}}\label{line:factor}
    }
    \BoxingPhase{
      \Return $\reduce{\shortZ{c}{\left[\,\widehat{G}^*\ G_K\right]}}{\,\girard}{\,\Zorder}$\;\label{line:box}
    }
\end{algorithm}

An overview of RAZeR is given in \cref{alg:razer}, consisting of two phases:
In the \emph{absorption phase}, it absorbs the generators with an expected positive absorption benefit,
which can be computed using  \cref{def:boxcost} and \cref{def:abscost}:
\begin{definition}[Absorption benefit]
 \label[definition]{def:benefit}
For a linear map $M\in\R^{m\times n}$, a basis $G^*\in\R^{n\times n}$, and a generator $g\in\R^n$ with $v=(G^*)^{-1}g$, the absorption benefit of $g$ is given by:
\begin{equation*}
  b(v) := b_M(v)=\boxcost_M(g)-\abscost_M(g)=\norm{\abs{M}\abs{g}}_1-\norm{\abs{MG^*}\abs{v}}_1.
\end{equation*}%
\end{definition}
Afterward, the \emph{boxing phase} encloses all remaining generators using~\eqref{eq:rad} to obtain the desired order.

Please note that the computed absorption benefit depends on the propagation to the next time step, i.e., on the matrix $M$.
For classic reachability analysis, this is computed directly during linearization (\cref{alg:reach}, line~\ref{alg-line:compute-map}).
However, as order reduction is typically applied after the propagation as a post-processing step (line~\ref{alg-line:reduce}), we argue that it should rather be seen as a pre-processing step to prepare a given zonotope for the upcoming propagation.

In settings where we do not have propagation information,
such as runtime monitoring under uncertainty~\cite{DBLP:conf/rv/FinkbeinerFKK24},
we can only estimate the costs of the upcoming propagation:
We approximate the absorption benefit using the immediate cost with the identity map, \(\abscost_{I_n}(g)\). 
However, an equivalent approach to estimating the boxing cost would presume that boxing incurs no wrapping cost, as  \(\boxcost_{I_n}(g)=0\). To be able to account for the error boxing incurs over several reduction steps, we estimate the boxing cost as \(\aprxboxcost(g)=\tau\|g\|_1\), where \(\tau\geq0\) controls the preference for absorption and can be calibrated as shown in \cref{fig:pzr}. Substituting these costs into \cref{def:benefit} yields
\begin{equation}
  b(g) := b_\tau(g)=  \aprxboxcost(g) - \abscost_{I_n}(g) = \tau\norm{g}_1 -\abscost_{I_n}(g).
  \label{eq:eta}
\end{equation}
At $\tau=0$, \method{} would only absorb when it would be lossless and behave similarly to \girard{}, while at $\tau\rightarrow\infty$, \method{} would absorb all surplus generators.
Crucially, \cref{alg:razer} computes a sound outer approximation independent of the choice of $M$ and $\tau$.
\begin{proposition}[\method{} is sound]
    Given $\Z=\shortZ{c}{G}\subset\R^n$, a maximal order $\Zorder\geq 1$, and a matrix $M\in\R^{n\times n}$, \Cref{alg:razer} computes a zonotope $\widehat{\Z}\supseteq\Z$ with order at most $\Zorder$. 
    \begin{proof}
        The outer approximation is given by replicating the absorption of \scott{} for the generators with a positive absorption benefit, and by the box enclosure of \girard{} for the remaining generators. The order is ensured by the application of \girard{} on the zonotope with the absorbed generators (line~\ref{line:box}). Thus, the choice of $M$ and $\tau$ only affect tightness.
    \end{proof}
\end{proposition}

We can further show that \method{} 's ability to combine absorption and boxing of surplus vectors yields a better solution than if we treated all generators in the same manner. 

\begin{proposition}[Mixing is better]
\label{prop:local-opt}
Let $G^*$ be a fixed invertible basis and $M$ a linear map. For the surplus generators $g_i$ of $G_S$, let $v_i=(G^*)^{-1}g_i$. Then, selecting absorption when $b_M(v_i)>0$ and boxing otherwise yields
\begin{equation}
\label{eq:local-dominance}
\sum_{g_i}\min\{\boxcost_M(g_i),\abscost_M(g_i)\}
\leq
\min\!\left\{\sum_{g_i}\boxcost_M(g_i),\,
              \sum_{g_i}\abscost_M(g_i)\right\}.
\end{equation}
\begin{proof}
By \cref{def:benefit}, $b_M(v_i)=\boxcost_M(g_i)-\abscost_M(g_i)$, so the selected operation incurs $\min\{\boxcost_M(g_i),\abscost_M(g_i)\}$. The claim follows by triangle inequality.
\end{proof}
\end{proposition}

\section{Evaluation}\label{sec:eval}

We evaluate \method{} on eleven scenarios from nonlinear reachability from ARCH-COMP~\cite{ARCH_COMP26_Category_Report} and previous literature on adaptive reachability~\cite{wetzlinger2022adaptive}{}.
All experiments are run in MATLAB R2026a with CORA v2026.1.0~\cite{althoff2025cora} on a laptop with an Intel Core Ultra 7 155U and 32 GB of RAM under Fedora Linux 44.

\subsection{Main Results}
We compare the performance of \method{} against individual reduction methods from \cref{sec:background} using orders $\Zorder\in\{3,5,10,20\}$. 
For all experiments, we show the relative tightness of \method{} against the respective method in the final time step computed by:  
\begin{equation}
    \label{eq:eval-metric}
    \rad(\nicefrac{\mathtt{method}}{\method{}}) = \frac{\norm{\rad(\Z_{\mathtt{method}})}_1}{\norm{\rad(\Z_{\method{}})}_1}.
\end{equation}

\begin{figure}[h]
\centering
\input{generated/tikz_summary_tradeoff}
\caption{\method{} compared to the best of \girard{}, \combastel{}, \scott{} and \pca{} per configuration.}
\label{fig:summary}
\end{figure}

\begin{figure}[h]
\centering
\begin{tikzpicture}[inner sep=0pt, line cap=butt]
\definecolor{scott}{HTML}{3B8BC2}
\definecolor{pca}{HTML}{D55E00}
\definecolor{combastel}{HTML}{D4B300}
\definecolor{girard}{HTML}{E69F00}
\definecolor{horizon_dt}{HTML}{009E73}
\useasboundingbox (0,0) rectangle (8.458,6.096);
\draw[scott, line width=1.2pt, line join=miter, dash pattern=on 1.2pt off 1.2pt] (1.057,1.657) -- (1.221,1.657) -- (1.221,1.743) -- (4.068,1.743) -- (4.068,1.83) -- (4.294,1.83) -- (4.294,1.917) -- (4.802,1.917) -- (4.802,2.003) -- (5.113,2.003) -- (5.113,2.09) -- (6.978,2.09);
\node[anchor=west, font=\fontsize{6.38}{6.364}\selectfont, align=left, text=scott] at (7.103,2.121) {\scott{}\\ completes 9};
\draw[pca, line width=0.9pt, line join=miter, dash pattern=on 3.6pt off 1.35pt] (1.057,1.484) -- (1.079,1.484) -- (1.079,1.57) -- (1.116,1.57) -- (1.116,1.657) -- (1.162,1.657) -- (1.162,1.743) -- (1.269,1.743) -- (1.269,1.83) -- (1.283,1.83) -- (1.283,1.917) -- (1.522,1.917) -- (1.522,2.003) -- (1.624,2.003) -- (1.624,2.09) -- (1.686,2.09) -- (1.686,2.176) -- (1.84,2.176) -- (1.84,2.263) -- (1.937,2.263) -- (1.937,2.35) -- (2.775,2.35) -- (2.775,2.436) -- (2.795,2.436) -- (2.795,2.523) -- (3.241,2.523) -- (3.241,2.609) -- (4.49,2.609) -- (4.49,2.696) -- (5.766,2.696) -- (5.766,2.782) -- (6.022,2.782) -- (6.022,2.869) -- (6.978,2.869);
\node[anchor=west, font=\fontsize{6.38}{6.364}\selectfont, align=left, text=pca] at (7.103,2.883) {\pca{}\\ completes 18};
\draw[combastel, line width=1pt, line join=miter, dash pattern=on 6pt off 1.5pt on 1pt off 1.5pt] (1.057,1.743) -- (1.069,1.743) -- (1.069,1.83) -- (1.074,1.83) -- (1.074,1.917) -- (1.075,1.917) -- (1.075,2.003) -- (1.09,2.003) -- (1.09,2.09) -- (1.096,2.09) -- (1.096,2.176) -- (1.137,2.176) -- (1.137,2.263) -- (1.143,2.263) -- (1.143,2.35) -- (1.162,2.35) -- (1.162,2.436) -- (1.162,2.436) -- (1.162,2.523) -- (1.182,2.523) -- (1.182,2.609) -- (1.187,2.609) -- (1.187,2.696) -- (1.196,2.696) -- (1.196,2.782) -- (1.199,2.782) -- (1.199,2.869) -- (1.241,2.869) -- (1.241,2.956) -- (1.343,2.956) -- (1.343,3.042) -- (1.347,3.042) -- (1.347,3.129) -- (1.347,3.129) -- (1.347,3.215) -- (1.347,3.215) -- (1.347,3.302) -- (1.382,3.302) -- (1.382,3.389) -- (1.411,3.389) -- (1.411,3.475) -- (1.502,3.475) -- (1.502,3.562) -- (1.516,3.562) -- (1.516,3.648) -- (1.559,3.648) -- (1.559,3.735) -- (1.607,3.735) -- (1.607,3.822) -- (1.618,3.822) -- (1.618,3.908) -- (1.71,3.908) -- (1.71,3.995) -- (1.852,3.995) -- (1.852,4.081) -- (2.579,4.081) -- (2.579,4.168) -- (3.435,4.168) -- (3.435,4.255) -- (3.71,4.255) -- (3.71,4.341) -- (3.817,4.341) -- (3.817,4.428) -- (3.853,4.428) -- (3.853,4.514) -- (4.017,4.514) -- (4.017,4.601) -- (4.052,4.601) -- (4.052,4.687) -- (5.096,4.687) -- (5.096,4.774) -- (6.978,4.774);
\node[anchor=west, font=\fontsize{6.38}{6.364}\selectfont, align=left, text=combastel] at (7.103,3.912) {\combastel{}\\ completes 39};
\draw[girard, line width=1pt, line join=miter] (1.057,2.436) -- (1.07,2.436) -- (1.07,2.523) -- (1.078,2.523) -- (1.078,2.609) -- (1.078,2.609) -- (1.078,2.696) -- (1.095,2.696) -- (1.095,2.782) -- (1.097,2.782) -- (1.097,2.869) -- (1.12,2.869) -- (1.12,2.956) -- (1.125,2.956) -- (1.125,3.042) -- (1.126,3.042) -- (1.126,3.129) -- (1.169,3.129) -- (1.169,3.215) -- (1.177,3.215) -- (1.177,3.302) -- (1.218,3.302) -- (1.218,3.389) -- (1.272,3.389) -- (1.272,3.475) -- (1.278,3.475) -- (1.278,3.562) -- (1.301,3.562) -- (1.301,3.648) -- (1.312,3.648) -- (1.312,3.735) -- (1.313,3.735) -- (1.313,3.822) -- (1.636,3.822) -- (1.636,3.908) -- (1.712,3.908) -- (1.712,3.995) -- (1.836,3.995) -- (1.836,4.081) -- (1.838,4.081) -- (1.838,4.168) -- (1.894,4.168) -- (1.894,4.255) -- (1.933,4.255) -- (1.933,4.341) -- (1.975,4.341) -- (1.975,4.428) -- (2.006,4.428) -- (2.006,4.514) -- (2.196,4.514) -- (2.196,4.601) -- (2.886,4.601) -- (2.886,4.687) -- (3.19,4.687) -- (3.19,4.774) -- (3.773,4.774) -- (3.773,4.861) -- (4.62,4.861) -- (4.62,4.947) -- (6.978,4.947);
\node[anchor=west, font=\fontsize{6.38}{6.364}\selectfont, align=left, text=girard] at (7.103,4.712) {\girard{}\\ completes 43};
\node[anchor=east, font=\fontsize{6.05}{7.305}\selectfont, text=girard] at (0.951,2.436) {27\,\%};
\draw[horizon_dt, line width=1.5pt, line join=miter] (1.057,4.514) -- (1.071,4.514) -- (1.071,4.601) -- (1.519,4.601) -- (1.519,4.687) -- (1.659,4.687) -- (1.659,4.774) -- (1.686,4.774) -- (1.686,4.861) -- (2.107,4.861) -- (2.107,4.947) -- (5.559,4.947) -- (5.559,5.034) -- (6.236,5.034) -- (6.236,5.12) -- (6.689,5.12) -- (6.689,5.207) -- (6.978,5.207);
\node[anchor=west, font=\fontsize{6.38}{6.364}\selectfont, align=left, text=horizon_dt] at (7.103,5.207) {\method{}\\ completes 44};
\node[anchor=east, font=\fontsize{6.05}{7.305}\selectfont, text=horizon_dt] at (0.951,4.514) {82\,\%};
\draw[line width=0.5pt, line cap=rect] (1.057,1.397) -- (6.978,1.397);
\draw[line width=0.5pt] (1.057,1.397) -- (1.057,1.326);
\node[anchor=north, font=\fontsize{7.15}{8.634}\selectfont, text height=.7em, text depth=.2em] at (1.057,1.203) {1};
\draw[line width=0.5pt] (2.1,1.397) -- (2.1,1.326);
\node[anchor=north, font=\fontsize{7.15}{8.634}\selectfont, text height=.7em, text depth=.2em] at (2.1,1.203) {1.5};
\draw[line width=0.5pt] (2.84,1.397) -- (2.84,1.326);
\node[anchor=north, font=\fontsize{7.15}{8.634}\selectfont, text height=.7em, text depth=.2em] at (2.84,1.203) {2};
\draw[line width=0.5pt] (3.882,1.397) -- (3.882,1.326);
\node[anchor=north, font=\fontsize{7.15}{8.634}\selectfont, text height=.7em, text depth=.2em] at (3.882,1.203) {3};
\draw[line width=0.5pt] (5.196,1.397) -- (5.196,1.326);
\node[anchor=north, font=\fontsize{7.15}{8.634}\selectfont, text height=.7em, text depth=.2em] at (5.196,1.203) {5};
\draw[line width=0.5pt] (6.978,1.397) -- (6.978,1.326);
\node[anchor=north, font=\fontsize{7.15}{8.634}\selectfont, text height=.7em, text depth=.2em] at (6.978,1.203) {10};
\draw[line width=0.5pt] (1.057,1.397) -- (1.057,5.207);
\draw[line width=0.5pt] (1.057,1.397) -- (0.987,1.397);
\node[anchor=east, font=\fontsize{7.15}{8.634}\selectfont] at (0.669,1.397) {0};
\draw[line width=0.5pt] (1.057,2.35) -- (0.987,2.35);
\draw[line width=0.5pt] (1.057,3.302) -- (0.987,3.302);
\node[anchor=east, font=\fontsize{7.15}{8.634}\selectfont] (yt2) at (0.669,3.302) {0.5};
\draw[line width=0.5pt] (1.057,4.255) -- (0.987,4.255);
\draw[line width=0.5pt] (1.057,5.207) -- (0.987,5.207);
\node[anchor=east, font=\fontsize{7.15}{8.634}\selectfont] at (0.669,5.207) {1};
\node[anchor=north, font=\fontsize{7.7}{9.298}\selectfont] at (4.018,0.941) {within a factor $\theta$ of the tightest of the five methods};
\coordinate (yl) at (yt2.west |- 0,3.302);
\node[anchor=south, rotate=90, font=\fontsize{7.7}{9.24}\selectfont] at ([xshift=-1pt]yl) {share of the 44 configurations};
\end{tikzpicture}%
\caption{Share of completed runs of each method having its tightness within $\theta$ times the radius of the best method.}
\label{fig:profile}
\end{figure}
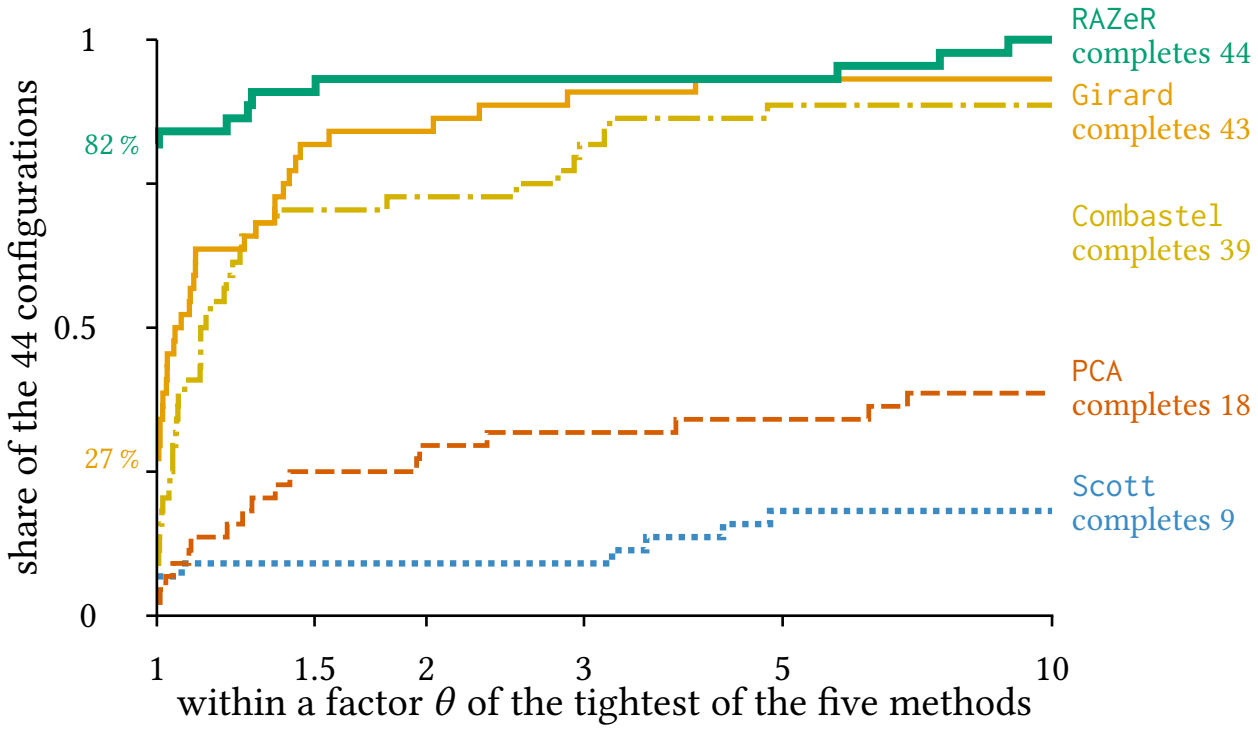

\method{} obtains substantially tighter results in almost all benchmarks (\cref{fig:dots})---even if it is compared to the best competing method of each benchmark (\cref{fig:summary}), which is of course not known in advance.
\cref{fig:summary} also shows that the runtime overhead for \method{} against the competing method is on average only $10\%$, 
while automatically tuning the reduction method and thus obtaining tighter results.
This can also be seen in \cref{fig:profile}, where the radius of \method{} is within the same order of magnitude as the tightest other method, and even within a factor $\theta\le 1.5$ in most cases. 
We have only observed worse behavior on the ROBE and JET benchmarks, which we attribute to system dynamics in these scenarios that lead to inaccurate next-map predictions and, therefore, to suboptimal choices of absorption or boxing, resulting in looser zonotopes.

\subsection{Ablation Studies}

\begin{figure}[h]
\centering
\input{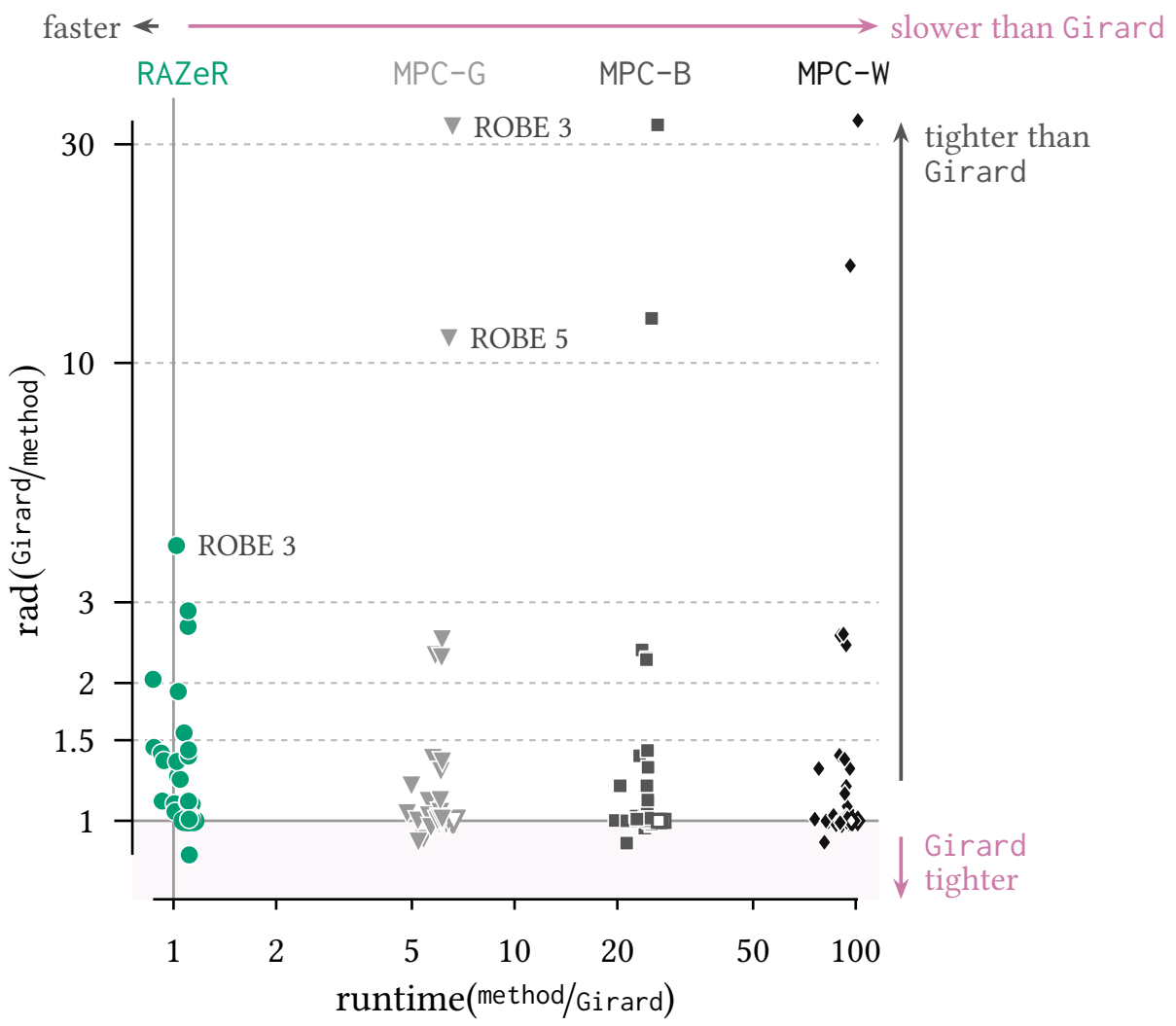}
\caption{Comparison of \method{} and the model predictive controllers, both normalized to \girard{}.}
\label{fig:mpc}
\end{figure}

\paragraph{\method{} vs model predictive control}
\method{} is motivated by model predictive control, but does not come with the computational overhead that usually comes from model predictive control. 
We adapt the existing model predictive controller for zonotope reduction~\cite{krsmanovic2026predictive}  to better suit the ARCH-COMP setting, 
and limit it to selecting any of the four reduction methods presented in \cref{sec:background}. 
We run several variants of these controllers: MPC-G (greedy search, $H=1$), MPC-B ($H=3$, $W=2$), MPC-W ($H=5$, $W=4$),
with hyperparameter names as defined in \cite{krsmanovic2026predictive}.
The hyperparameter $H$ effectively counts the number of future time steps considered for the prediction, 
such that we can say that \method{} uses $H=1$ as it also greedily looks one time step into the future.
However, as shown in \cref{fig:mpc}, \method{} matches or even outperforms these model predictive controllers in many scenarios at a fraction of the computational cost; e.g., it is, on average, $5\times$ faster to compute than the next-fastest controller, MPC-G, and $100\times$ faster than MPC-W.

\paragraph{Impact of $\tau$}

\begin{figure}[h]
\centering
\resizebox{0.92\columnwidth}{!}{\input{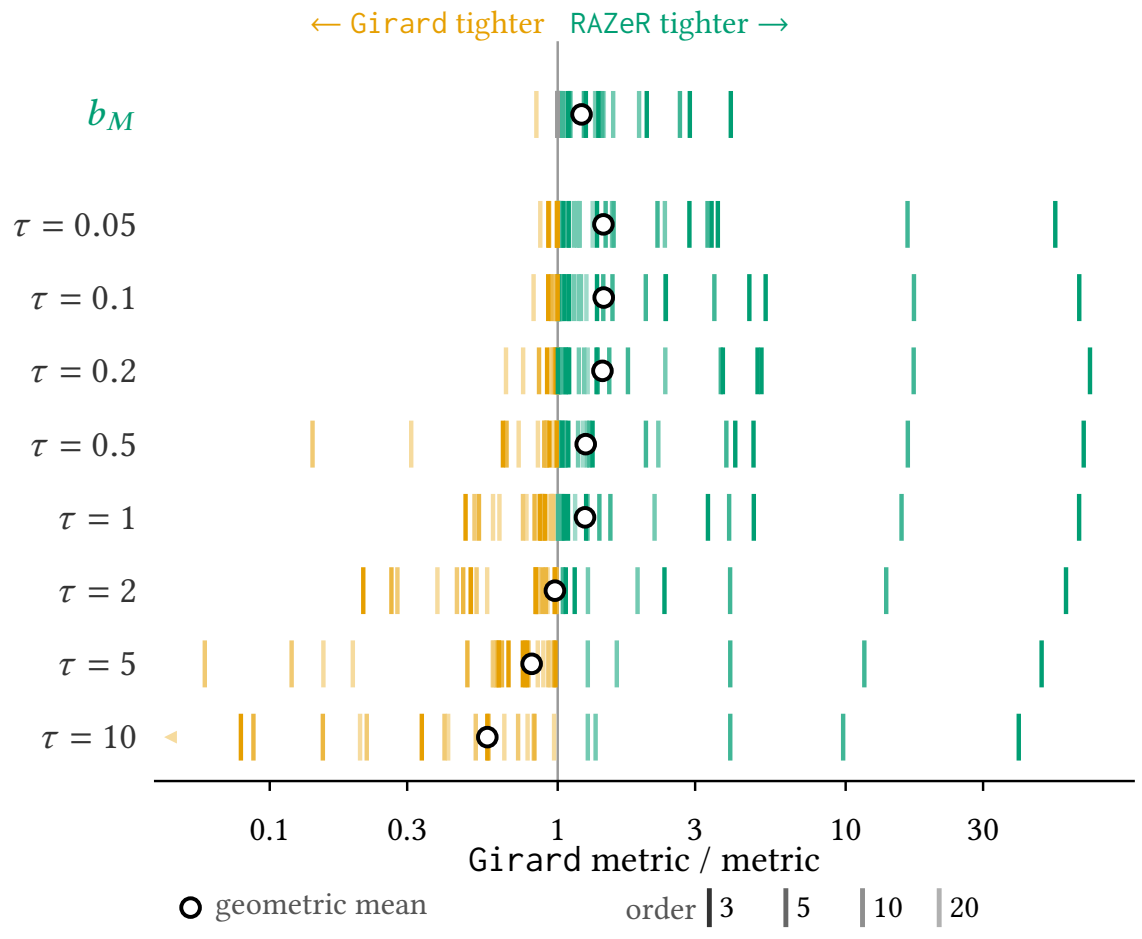}}
\caption{Ablation on absorption threshold $\tau$.}
\label{fig:ablation}
\end{figure}

The threshold parameter $\tau$ controls the inflation of absorbing methods~\eqref{eq:eta}, and we perform a sweep of its values in \cref{fig:ablation}. 
While too much absorption ($\tau\ge 2$) has a negative influence, as we have also seen with \scott{} above, 
we found that a moderate balance ($\tau=0.1$) between the absorption and boxing obtains the best result.

\begin{figure}[h]
\centering
\resizebox{0.92\columnwidth}{!}{
\begin{tikzpicture}[inner sep=0pt, line cap=butt]
\definecolor{razer}{HTML}{009E73}
\definecolor{razerf}{HTML}{66A61E}
\definecolor{tie}{HTML}{8C8C8C}
\definecolor{diag}{HTML}{777777}
\definecolor{one}{HTML}{AAAAAA}
\definecolor{note}{HTML}{444444}
\definecolor{lead}{HTML}{888888}
\definecolor{keylabel}{HTML}{555555}
\definecolor{key}{HTML}{555555}
\useasboundingbox (0,0) rectangle (8.458,5.694);
\begin{scope}\clip (1.1,1.27) rectangle (8.204,5.491);
\draw[fill=razer!7!white, draw=none] (-2.997,-2.755) -- (9.773,10.015) -- (-2.997,10.015) -- cycle;
\draw[fill=razerf!7!white, draw=none] (-2.997,-2.755) -- (9.773,10.015) -- (9.773,-2.755) -- cycle;
\fill[black, fill opacity=.03] (1.1,1.27) rectangle (1.783,5.491);
\fill[black, fill opacity=.03] (1.1,1.27) rectangle (8.204,2.025);
\draw[diag, line width=.5pt] (-2.997,-2.755) -- (9.773,10.015);
\end{scope}
\draw[one, line width=.4pt] (1.783,1.27) -- (1.783,5.491);
\draw[one, line width=.4pt] (1.1,2.025) -- (8.204,2.025);
\filldraw[fill=razer!25!white, draw=white, line width=0.3pt] (1.838,3.105) circle (1.65pt);
\filldraw[fill=razer!25!white, draw=white, line width=0.3pt] (1.849,2.166) circle (1.65pt);
\filldraw[fill=razer!25!white, draw=white, line width=0.3pt] (2.629,2.916) circle (1.65pt);
\filldraw[fill=razerf!25!white, draw=white, line width=0.3pt] (1.81,2.051) circle (1.65pt);
\filldraw[fill=white, draw=tie!25!white, line width=0.6pt] (1.783,2.025) circle (1.5pt);
\filldraw[fill=razer!25!white, draw=white, line width=0.3pt] (1.815,2.073) circle (1.65pt);
\filldraw[fill=white, draw=tie!25!white, line width=0.6pt] (1.783,2.025) circle (1.5pt);
\filldraw[fill=razerf!25!white, draw=white, line width=0.3pt] (1.292,1.491) circle (1.65pt);
\filldraw[fill=razerf!25!white, draw=white, line width=0.3pt] (1.839,2.075) circle (1.65pt);
\filldraw[fill=white, draw=tie!25!white, line width=0.6pt] (1.783,2.025) circle (1.5pt);
\filldraw[fill=razer!25!white, draw=white, line width=0.3pt] (1.807,2.052) circle (1.65pt);
\filldraw[fill=razer!45!white, draw=white, line width=0.3pt] (1.897,3.294) circle (1.65pt);
\filldraw[fill=razer!45!white, draw=white, line width=0.3pt] (2.245,2.93) circle (1.65pt);
\filldraw[fill=razerf!45!white, draw=white, line width=0.3pt] (4.025,3.785) circle (1.65pt);
\filldraw[fill=razerf!45!white, draw=white, line width=0.3pt] (1.803,2.041) circle (1.65pt);
\filldraw[fill=white, draw=tie!45!white, line width=0.6pt] (1.783,2.025) circle (1.5pt);
\filldraw[fill=razer!45!white, draw=white, line width=0.3pt] (1.83,2.105) circle (1.65pt);
\filldraw[fill=white, draw=tie!45!white, line width=0.6pt] (1.783,2.025) circle (1.5pt);
\filldraw[fill=razerf!45!white, draw=white, line width=0.3pt] (1.912,2.058) circle (1.65pt);
\filldraw[fill=razerf!45!white, draw=white, line width=0.3pt] (1.872,2.112) circle (1.65pt);
\filldraw[fill=white, draw=tie!45!white, line width=0.6pt] (1.783,2.025) circle (1.5pt);
\filldraw[fill=razerf!45!white, draw=white, line width=0.3pt] (2.175,2.34) circle (1.65pt);
\filldraw[fill=razer!70!white, draw=white, line width=0.3pt] (1.804,2.343) circle (1.65pt);
\filldraw[fill=razer!70!white, draw=white, line width=0.3pt] (1.57,3.031) circle (1.65pt);
\filldraw[fill=razerf!70!white, draw=white, line width=0.3pt] (6.923,2.667) circle (1.65pt);
\filldraw[fill=razerf!70!white, draw=white, line width=0.3pt] (1.935,2.176) circle (1.65pt);
\filldraw[fill=white, draw=tie!70!white, line width=0.6pt] (1.783,2.025) circle (1.5pt);
\filldraw[fill=razer!70!white, draw=white, line width=0.3pt] (1.812,2.111) circle (1.65pt);
\filldraw[fill=white, draw=tie!70!white, line width=0.6pt] (1.783,2.025) circle (1.5pt);
\filldraw[fill=razerf!70!white, draw=white, line width=0.3pt] (4.239,4.468) circle (1.65pt);
\filldraw[fill=razer!70!white, draw=white, line width=0.3pt] (2.054,2.327) circle (1.65pt);
\filldraw[fill=white, draw=tie!70!white, line width=0.6pt] (1.783,2.025) circle (1.5pt);
\filldraw[fill=razerf!70!white, draw=white, line width=0.3pt] (2.94,3.074) circle (1.65pt);
\filldraw[fill=razerf, draw=white, line width=0.3pt] (2.625,2.714) circle (1.65pt);
\filldraw[fill=razer, draw=white, line width=0.3pt] (3.324,3.921) circle (1.65pt);
\filldraw[fill=razerf, draw=white, line width=0.3pt] (8.019,5.195) circle (1.65pt);
\filldraw[fill=razer, draw=white, line width=0.3pt] (2.01,2.3) circle (1.65pt);
\filldraw[fill=white, draw=tie, line width=0.6pt] (1.783,2.025) circle (1.5pt);
\filldraw[fill=razer, draw=white, line width=0.3pt] (1.767,2.293) circle (1.65pt);
\filldraw[fill=white, draw=tie, line width=0.6pt] (1.783,2.025) circle (1.5pt);
\filldraw[fill=razer, draw=white, line width=0.3pt] (2.646,2.991) circle (1.65pt);
\filldraw[fill=white, draw=tie, line width=0.6pt] (1.783,2.025) circle (1.5pt);
\filldraw[fill=razer, draw=white, line width=0.3pt] (3.945,4.618) circle (1.65pt);
\draw[lead, line width=.4pt] (1.57,3.109) -- (1.57,4.39);
\node[anchor=south west, font=\fontsize{6.05}{7.305}\selectfont, text=note] at (1.499,4.426) {LALO-W01 5};
\node[anchor=east, font=\fontsize{6.05}{7.305}\selectfont, text=note] at (7.878,5.195) {ROBE 3};
\node[anchor=east, font=\fontsize{6.05}{7.305}\selectfont, text=note] at (6.782,2.667) {ROBE 5};
\node[anchor=west, font=\fontsize{6.05}{7.305}\selectfont, text=note] at (1.433,1.491) {JET 20};
\node[anchor=north west, font=\fontsize{6.6}{7.97}\selectfont, text=razer] at (1.242,5.322) {\method{} tighter (17)};
\node[anchor=south east, font=\fontsize{6.6}{7.97}\selectfont, text=razerf] at (8.062,1.439) {\mth{RAZeR-F} tighter (14)};
\draw[line width=0.5pt] (1.292,1.27) -- (8.019,1.27);
\draw[line width=0.5pt] (1.783,1.27) -- (1.783,1.199);
\node[anchor=north, font=\fontsize{7.15}{8.634}\selectfont, text height=.7em, text depth=.2em] at (1.783,1.076) {1};
\draw[line width=0.5pt] (3.64,1.27) -- (3.64,1.199);
\node[anchor=north, font=\fontsize{7.15}{8.634}\selectfont, text height=.7em, text depth=.2em] at (3.64,1.076) {2};
\draw[line width=0.5pt] (4.957,1.27) -- (4.957,1.199);
\node[anchor=north, font=\fontsize{7.15}{8.634}\selectfont, text height=.7em, text depth=.2em] at (4.957,1.076) {4};
\draw[line width=0.5pt] (6.265,1.27) -- (6.265,1.199);
\node[anchor=north, font=\fontsize{7.15}{8.634}\selectfont, text height=.7em, text depth=.2em] at (6.265,1.076) {10};
\draw[line width=0.5pt] (7.923,1.27) -- (7.923,1.199);
\node[anchor=north, font=\fontsize{7.15}{8.634}\selectfont, text height=.7em, text depth=.2em] at (7.923,1.076) {50};
\draw[line width=0.5pt] (1.1,1.491) -- (1.1,5.195);
\draw[line width=0.5pt] (1.1,2.025) -- (0.976,2.025);
\node[anchor=east, font=\fontsize{7.15}{8.634}\selectfont] (yt0) at (0.853,2.025) {1};
\draw[line width=0.5pt] (1.1,3.882) -- (0.976,3.882);
\node[anchor=east, font=\fontsize{7.15}{8.634}\selectfont] at (0.853,3.882) {2};
\draw[line width=0.5pt] (1.1,5.199) -- (0.976,5.199);
\node[anchor=east, font=\fontsize{7.15}{8.634}\selectfont] at (0.853,5.199) {4};
\node[anchor=north, font=\fontsize{7.7}{9.298}\selectfont] at (4.652,0.814) {$\rad(\nicefrac{\girard{}}{\method{}\texttt{-F}})$};
\coordinate (yl) at (yt0.west |- 0,3.38);
\node[anchor=south, rotate=90, font=\fontsize{7.7}{9.24}\selectfont] at ([xshift=-1pt]yl) {$\rad(\nicefrac{\girard{}}{\method{}})$};
\node[anchor=east, font=\fontsize{6.6}{7.97}\selectfont, text=keylabel] at (4.1,0.152) {Order $\Zorder$:};
\filldraw[fill=key, draw=white, line width=0.3pt] (4.356,0.152) circle (1.65pt);
\node[anchor=west, font=\fontsize{6.6}{7.97}\selectfont] at (4.483,0.152) {3};
\filldraw[fill=key!70!white, draw=white, line width=0.3pt] (4.99,0.152) circle (1.65pt);
\node[anchor=west, font=\fontsize{6.6}{7.97}\selectfont] at (5.117,0.152) {5};
\filldraw[fill=key!45!white, draw=white, line width=0.3pt] (5.625,0.152) circle (1.65pt);
\node[anchor=west, font=\fontsize{6.6}{7.97}\selectfont] at (5.752,0.152) {10};
\filldraw[fill=key!25!white, draw=white, line width=0.3pt] (6.259,0.152) circle (1.65pt);
\node[anchor=west, font=\fontsize{6.6}{7.97}\selectfont] at (6.386,0.152) {20};
\end{tikzpicture}%
}
\caption{Deciding per generator (\method{}) or per matrix (\mth{RAZeR-F}), both against \girard{}.}
\label{fig:granularity}
\end{figure}
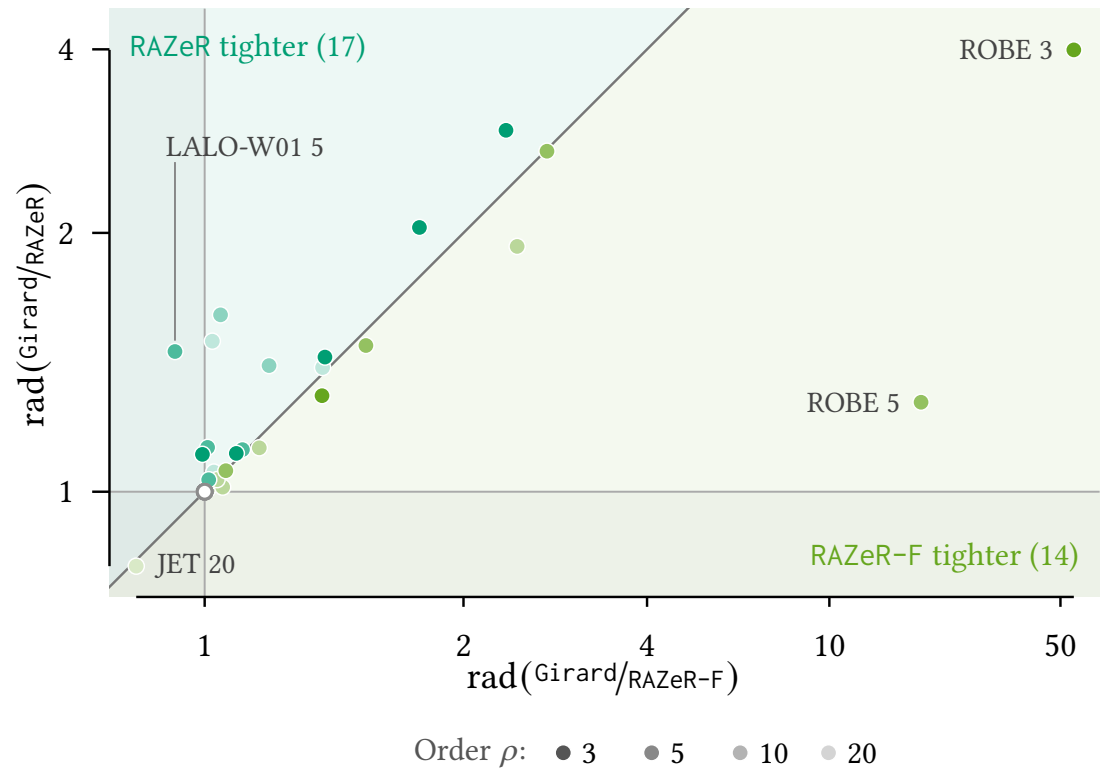

\paragraph{Granularity vs. coarseness}
\method{} operates at the generator level, i.e., it decides for each generator whether it should be absorbed or boxed.
To determine if this granularity is necessary, we compare it to a variant, \method{}-F, 
which decides on the reduction for the full zonotope using the cumulative absorption benefit (\cref{def:benefit}). 
The results in \cref{fig:granularity} show that \method{} yields tighter zonotopes compared with \method{}-F for most benchmarks,
while \method{}-F is substantially tighter for the two ROBE benchmarks.

\section{Limitations}
\method{} is well-suited to sequential zonotope-reduction applications such as reachability analysis, runtime monitoring under uncertainty, or neural network verification. 
Its effectiveness depends on the linearization of the system dynamics model, or, when this is unavailable, on the choice of a heuristic threshold that must be tuned.
In static settings, where there is no temporal propagation of zonotopes, its advantage is limited and comparable to that of existing methods. 
However, we can push the outer approximation induced by the reduction away from an unsafe specification boundary using \method{}:
For example, let the specification be given by a polytope $\mathcal{P} = \{x\in\R^n\ |\ Ax\le b\}$ for some $A\in\R^{p\times n}$ and $b\in\R^p$; then, choosing $M=A$ in \cref{def:benefit} reduces the outer approximation towards the decision boundary as boxing error in the directions specified in $A$ is minimized.
While \method{} performs well across the ARCH-COMP settings, it does not uniformly outperform all existing methods and incurs additional computational overhead.

\section{Conclusion}
We presented \method{}, a new zonotope reduction method that dynamically balances generator absorption and boxing based on the properties of the given zonotope and its upcoming propagation. 
We argue that zonotope reductions are not just a post-processing step, but also a pre-processing step that shapes how the zonotope evolves with the system dynamics. 
We use this perspective to derive a computationally cheap heuristic operating at the generator level, thereby deciding whether to box or absorb each surplus generator. 
This approach combines the complementary strengths of existing reduction methods without requiring an explicit predictive optimization loop.
Our evaluation of nonlinear reachability benchmarks demonstrates that \method{} produces tighter enclosures than established reduction methods in most benchmarks, while preserving sound outer-approximative guarantees. 
\method{} matches or outperforms the results of a portfolio of existing reduction methods in 82\% of cases, with only moderate runtime overhead.


\bibliographystyle{ACM-Reference-Format}
\bibliography{sample-base}

@inproceedings{DBLP:journals/corr/cuttingcorners,
  title={Cutting Corners on Uncertainty: Zonotope Abstractions for Stream-based Runtime Monitoring},
  author={Finkbeiner, Bernd and Fr{\"a}nzle, Martin and Kohn, Florian and Kr{\"o}ger, Paul},
  booktitle={International Conference on Runtime Verification},
  year={2026}
}

@inproceedings{DBLP:conf/rv/FinkbeinerFKK24,
  author       = {Bernd Finkbeiner and
                  Martin Fr{\"{a}}nzle and
                  Florian Kohn and
                  Paul Kr{\"{o}}ger},
  title        = {Stream-Based Monitoring Under Measurement Noise},
  booktitle    = {International Conference on Runtime Verification},
  year         = {2024},
}

@inproceedings{kopetzki2017methods,
  title={Methods for order reduction of zonotopes},
  author={Kopetzki, Anna-Kathrin and Sch{\"u}rmann, Bastian and Althoff, Matthias},
  booktitle={IEEE Annual Conference on Decision and Control},
  year={2017},
}

@inproceedings{wetzlinger2022adaptive,
  title={Adaptive reachability algorithms for nonlinear systems using abstraction error analysis},
  author={Wetzlinger, Mark and Kulmburg, Adrian and Le Penven, Alexis and Althoff, Matthias},
  journal={Nonlinear Analysis: Hybrid Systems},
  year={2022},
}

@inproceedings{wetzlinger2020adaptive,
  title={Adaptive parameter tuning for reachability analysis of linear systems},
  author={Wetzlinger, Mark and Kochdumper, Niklas and Althoff, Matthias},
  booktitle={IEEE Annual Conference on Decision and Control},
  year={2020},
}

@inproceedings{DBLP:journals/automatica/scott,
  author       = {Joseph K. Scott and
                  Davide Martino Raimondo and
                  Giuseppe Roberto Marseglia and
                  Richard D. Braatz},
  title        = {Constrained zonotopes: {A} new tool for set-based estimation and fault
                  detection},
  journal      = {Automatica},
  year         = {2016},
}

@inproceedings{DBLP:conf/hybrid/Girard05,
  title={Reachability of uncertain linear systems using zonotopes},
  author={Girard, Antoine},
  booktitle={International Conference on Hybrid Systems: Computation and Control},
  year={2005}
}

@inproceedings{DBLP:conf/eucc/Combastel03,
  title={A state bounding observer based on zonotopes},
  author={Combastel, Christophe},
  booktitle={2003 European Control Conference},
  year={2003},
}

@inproceedings{krsmanovic2026predictive,
  title={Predictive Zonotope Reduction: Precise Runtime Monitoring under Uncertainty},
  author={Krsmanovic, Vladimir and Kohn, Florian and Finkbeiner, Bernd and Simovic, Milan},
  journal={arXiv:2609.03699},
  year={2026}
}

@phdthesis{althoff2010reachability,
  title={Reachability analysis and its application to the safety assessment of autonomous cars},
  author={Althoff, Matthias},
  year={2010},
  school={Technische Universit{\"a}t M{\"u}nchen}
}

@article{althoff2025cora,
  title={CORA manual},
  author={Althoff, Matthias and Kochdumper, Niklas and Ladner, Tobias and Perschl, Maximilian and Wetzlinger, Mark},
  journal={Technical University of Munich},
  year={2025}
}

@article{dyer1998complexity,
  title={On the complexity of computing mixed volumes},
  author={Dyer, Martin and Gritzmann, Peter and Hufnagel, Alexander},
  journal={SIAM Journal on Computing},
  year={1998},
}

@article{singh2018fast,
  title={Fast and effective robustness certification},
  author={Singh, Gagandeep and Gehr, Timon and Mirman, Matthew and P{\"u}schel, Markus and Vechev, Martin},
  journal={NeurIPS},
  year={2018}
}

@inproceedings{sadraddini2019linear,
  title={Linear encodings for polytope containment problems},
  author={Sadraddini, Sadra and Tedrake, Russ},
  booktitle={IEEE Annual Conference on Decision and Control},
  year={2019}
}

@inproceedings{ARCH_COMP26_Category_Report,
  author    = {Luca Geretti and Julien Alexandre Dit Sandretto and Matthias Althoff and et al.},
  title     = {ARCH-COMP26 Category Report: Continuous and Hybrid Systems with Nonlinear Dynamics},
  booktitle    = {Workshop on Applied Verification for Continuous and Hybrid Systems},
  year      = {2026}
  }

\section*{Acknowledgements}
 This work was partially supported by the European Research Council (ERC) Grant HYPER (No.~101055412). Views and opinions expressed are, however, those of the authors only and do not necessarily reflect those of the European Union or the European Research Council Executive Agency. Neither the European Union nor the granting authority can be held responsible for them. This work was also supported by the project AL 1185/33-1 funded by the German Research Foundation (Deutsche Forschungsgemeinschaft, DFG). 
 

\appendix
\crefalias{section}{appendix}

\begin{table*}[b]
\centering\setlength{\tabcolsep}{3pt}
\caption{Absolute final metric values at fixed zonotope order $\rho$; smaller is tighter, and bold marks the smallest value of a row at the precision shown.  }
\label{tab:metric}
\resizebox{\textwidth}{!}{\begin{tabular}{llrrrrrrrrrrrrrr}
\toprule
 &  &  &  &  &  &  &  & \multicolumn{8}{c}{RAZeR, threshold $\tau$} \\
\cmidrule(lr){9-16}
Scenario & Metric & $\rho$ & Girard & Combastel & PCA & Scott & RAZeR & $0.05$ & $0.1$ & $0.2$ & $0.5$ & $1$ & $2$ & $5$ & $10$ \\
\midrule
LALO-W05 & $x_4$ & 3 & $1.69$ & F & F & F & $1.35$ & $0.469$ & $\mathbf{0.32}$ & $0.342$ & F & F & F & F & F \\
 &  & 5 & $0.203$ & $0.515$ & F & F & $0.183$ & $\mathbf{0.0913}$ & $0.1$ & $0.134$ & F & F & F & F & F \\
 &  & 10 & $0.0924$ & $0.188$ & F & F & $\mathbf{0.0593}$ & $0.0599$ & $0.0811$ & $0.0908$ & $0.656$ & F & F & F & F \\
 &  & 20 & $0.0605$ & $0.124$ & F & F & $\mathbf{0.0418}$ & $0.0508$ & $0.062$ & $0.0915$ & $0.195$ & F & F & F & F \\
\addlinespace[2pt]
LALO-W01 & $x_4$ & 3 & $0.0255$ & $0.0226$ & F & F & $0.0125$ & $7.46\!\times\!10^{-3}$ & $5.51\!\times\!10^{-3}$ & $\mathbf{5.01\!\times\!10^{-3}}$ & $6.17\!\times\!10^{-3}$ & $7.68\!\times\!10^{-3}$ & $0.0109$ & $0.0408$ & F \\
 &  & 5 & $6.48\!\times\!10^{-3}$ & $5.94\!\times\!10^{-3}$ & F & F & $4.61\!\times\!10^{-3}$ & $\mathbf{4.15\!\times\!10^{-3}}$ & $4.18\!\times\!10^{-3}$ & $4.71\!\times\!10^{-3}$ & $5.03\!\times\!10^{-3}$ & $4.64\!\times\!10^{-3}$ & $6.54\!\times\!10^{-3}$ & $0.0133$ & F \\
 &  & 10 & $4.38\!\times\!10^{-3}$ & $4.41\!\times\!10^{-3}$ & F & F & $\mathbf{3.24\!\times\!10^{-3}}$ & $3.85\!\times\!10^{-3}$ & $4.00\!\times\!10^{-3}$ & $4.13\!\times\!10^{-3}$ & $4.88\!\times\!10^{-3}$ & $5.79\!\times\!10^{-3}$ & $9.82\!\times\!10^{-3}$ & $0.0368$ & F \\
 &  & 20 & $3.25\!\times\!10^{-3}$ & $3.86\!\times\!10^{-3}$ & F & F & $\mathbf{3.11\!\times\!10^{-3}}$ & $3.74\!\times\!10^{-3}$ & $3.95\!\times\!10^{-3}$ & $4.29\!\times\!10^{-3}$ & $4.44\!\times\!10^{-3}$ & $5.46\!\times\!10^{-3}$ & $6.53\!\times\!10^{-3}$ & $0.0167$ & $0.419$ \\
\addlinespace[2pt]
ROBE & $2\,\omega$ & 3 & $0.0748$ & $2.50\!\times\!10^{-3}$ & $2.73\!\times\!10^{-3}$ & F & $0.0188$ & $1.40\!\times\!10^{-3}$ & $1.16\!\times\!10^{-3}$ & $\mathbf{1.06\!\times\!10^{-3}}$ & $1.12\!\times\!10^{-3}$ & $1.16\!\times\!10^{-3}$ & $1.29\!\times\!10^{-3}$ & $1.56\!\times\!10^{-3}$ & $1.88\!\times\!10^{-3}$ \\
 &  & 5 & $0.0175$ & $1.59\!\times\!10^{-3}$ & $3.13\!\times\!10^{-3}$ & $7.71\!\times\!10^{-3}$ & $0.0142$ & $1.07\!\times\!10^{-3}$ & $\mathbf{1.02\!\times\!10^{-3}}$ & $\mathbf{1.02\!\times\!10^{-3}}$ & $1.07\!\times\!10^{-3}$ & $1.12\!\times\!10^{-3}$ & $1.27\!\times\!10^{-3}$ & $1.51\!\times\!10^{-3}$ & $1.79\!\times\!10^{-3}$ \\
 &  & 10 & $2.36\!\times\!10^{-3}$ & $1.03\!\times\!10^{-3}$ & $1.05\!\times\!10^{-3}$ & $3.63\!\times\!10^{-3}$ & $1.23\!\times\!10^{-3}$ & $1.00\!\times\!10^{-3}$ & $\mathbf{9.98\!\times\!10^{-4}}$ & $\mathbf{9.98\!\times\!10^{-4}}$ & $1.06\!\times\!10^{-3}$ & $1.09\!\times\!10^{-3}$ & $1.25\!\times\!10^{-3}$ & $1.47\!\times\!10^{-3}$ & $1.74\!\times\!10^{-3}$ \\
 &  & 20 & $1.26\!\times\!10^{-3}$ & $\mathbf{9.32\!\times\!10^{-4}}$ & $9.40\!\times\!10^{-4}$ & $3.00\!\times\!10^{-3}$ & $9.37\!\times\!10^{-4}$ & $9.55\!\times\!10^{-4}$ & $1.00\!\times\!10^{-3}$ & $9.93\!\times\!10^{-4}$ & $1.03\!\times\!10^{-3}$ & $1.10\!\times\!10^{-3}$ & $1.21\!\times\!10^{-3}$ & $1.42\!\times\!10^{-3}$ & $1.61\!\times\!10^{-3}$ \\
\addlinespace[2pt]
PRDE & $x_3$ & 3 & $0.609$ & $0.634$ & F & F & $0.559$ & $0.559$ & $0.557$ & $\mathbf{0.556}$ & $0.56$ & $0.561$ & $0.572$ & $0.62$ & F \\
 &  & 5 & $0.582$ & $0.622$ & F & F & $0.556$ & $0.558$ & $0.556$ & $\mathbf{0.554}$ & $\mathbf{0.554}$ & $0.561$ & $0.559$ & $0.599$ & $0.702$ \\
 &  & 10 & $0.556$ & $0.619$ & $1.08$ & F & $0.553$ & $\mathbf{0.552}$ & $\mathbf{0.552}$ & $\mathbf{0.552}$ & $0.553$ & $0.553$ & $0.566$ & $0.599$ & $1.07$ \\
 &  & 20 & $0.554$ & $0.616$ & $0.702$ & F & $\mathbf{0.55}$ & $\mathbf{0.55}$ & $\mathbf{0.55}$ & $\mathbf{0.55}$ & $0.552$ & $0.555$ & $0.564$ & $0.591$ & $0.673$ \\
\addlinespace[2pt]
BIO & $x_1$ & 3 & $0.0187$ & $0.0194$ & F & F & $0.0187$ & $0.0188$ & $\mathbf{0.0186}$ & $0.019$ & $0.0199$ & $0.0206$ & $0.0374$ & F & F \\
 &  & 5 & $0.0177$ & $0.0183$ & F & F & $0.0177$ & $0.0176$ & $\mathbf{0.0174}$ & $0.0181$ & $0.0197$ & $0.0213$ & $0.0377$ & F & F \\
 &  & 10 & $\mathbf{0.0172}$ & $0.0179$ & F & F & $\mathbf{0.0172}$ & $\mathbf{0.0172}$ & $0.0174$ & $0.0179$ & $0.0188$ & $0.0194$ & $0.0329$ & $0.289$ & F \\
 &  & 20 & $0.0171$ & $0.0176$ & F & F & $0.0171$ & $\mathbf{0.017}$ & $0.0172$ & $0.018$ & $0.0186$ & $0.0192$ & $0.03$ & $0.111$ & F \\
\addlinespace[2pt]
BRUS & $x_1$ & 3 & $0.0995$ & $0.114$ & $0.114$ & F & $\mathbf{0.0915}$ & $0.095$ & $0.0942$ & $0.0934$ & $0.0959$ & $0.0948$ & $0.102$ & $0.147$ & $0.295$ \\
 &  & 5 & $0.0913$ & $0.106$ & $0.125$ & F & $0.0889$ & $0.0916$ & $0.0897$ & $\mathbf{0.0888}$ & $0.0893$ & $0.0911$ & $0.103$ & $0.148$ & $1.04$ \\
 &  & 10 & $0.0883$ & $0.102$ & $0.103$ & F & $0.0861$ & $\mathbf{0.086}$ & $0.0868$ & $0.0866$ & $0.0879$ & $0.0917$ & $0.103$ & $0.148$ & $0.218$ \\
 &  & 20 & $0.0868$ & $0.0982$ & $0.0929$ & F & $\mathbf{0.0856}$ & $0.0865$ & $0.086$ & $0.086$ & $0.0875$ & $0.0912$ & $0.102$ & $0.139$ & $0.209$ \\
\addlinespace[2pt]
GENE & $x_1$ & 3 & $\mathbf{0.111}$ & $0.113$ & F & F & $\mathbf{0.111}$ & $\mathbf{0.111}$ & $\mathbf{0.111}$ & $\mathbf{0.111}$ & $0.112$ & $0.129$ & $0.133$ & $0.142$ & $0.195$ \\
 &  & 5 & $0.11$ & $0.111$ & F & F & $0.11$ & $0.11$ & $0.111$ & $0.11$ & $\mathbf{0.108}$ & $0.126$ & $0.131$ & $0.143$ & $0.191$ \\
 &  & 10 & $\mathbf{0.108}$ & $0.109$ & $1.87\!\times\!10^{2}$ & $7.01$ & $\mathbf{0.108}$ & $\mathbf{0.108}$ & $\mathbf{0.108}$ & $0.109$ & $\mathbf{0.108}$ & $0.123$ & $0.129$ & $0.136$ & $0.148$ \\
 &  & 20 & $\mathbf{0.107}$ & $0.108$ & $0.667$ & $0.459$ & $\mathbf{0.107}$ & $\mathbf{0.107}$ & $0.108$ & $0.108$ & $\mathbf{0.107}$ & $0.112$ & $0.116$ & $0.125$ & $0.164$ \\
\addlinespace[2pt]
JET & $x_1$ & 3 & F & F & $0.105$ & $\mathbf{0.0276}$ & $0.159$ & $0.0489$ & $0.0433$ & $0.0399$ & $0.0305$ & $0.0286$ & $0.0278$ & $0.0277$ & $0.0277$ \\
 &  & 5 & $0.11$ & F & $0.0373$ & $\mathbf{0.0275}$ & $0.0414$ & $0.0332$ & $0.0314$ & $0.0299$ & $0.0286$ & $0.0279$ & $0.0277$ & $0.0277$ & $0.0277$ \\
 &  & 10 & $0.0352$ & $0.131$ & $0.0284$ & $\mathbf{0.0273}$ & $0.0349$ & $0.0302$ & $0.0293$ & $0.0285$ & $0.0279$ & $0.0277$ & $0.0277$ & $0.0277$ & $0.0277$ \\
 &  & 20 & $0.0269$ & $0.0738$ & $\mathbf{0.0252}$ & $0.0269$ & $0.0319$ & $0.0289$ & $0.0283$ & $0.028$ & $0.0277$ & $0.0277$ & $0.0277$ & $0.0277$ & $0.0277$ \\
\addlinespace[2pt]
LORE & $x_1$ & 3 & $0.371$ & $0.326$ & $0.627$ & F & $\mathbf{0.268}$ & $0.271$ & $0.271$ & $0.272$ & $0.281$ & $0.296$ & $0.324$ & $0.491$ & $4.67$ \\
 &  & 5 & $0.292$ & $0.296$ & F & F & $\mathbf{0.265}$ & $0.267$ & $0.267$ & $0.268$ & $0.274$ & $0.288$ & $0.321$ & $0.455$ & $1.91$ \\
 &  & 10 & $0.27$ & $0.283$ & F & F & $\mathbf{0.263}$ & $0.264$ & $0.265$ & $0.267$ & $0.272$ & $0.285$ & $0.317$ & $0.442$ & $1.25$ \\
 &  & 20 & $0.267$ & $0.276$ & $1.81$ & F & $\mathbf{0.263}$ & $0.264$ & $0.264$ & $0.266$ & $0.271$ & $0.284$ & $0.319$ & $0.449$ & $1.3$ \\
\addlinespace[2pt]
LOVO & $x_1$ & 3 & $\mathbf{0.138}$ & $0.14$ & F & F & $\mathbf{0.138}$ & $0.149$ & $0.149$ & $0.151$ & $0.214$ & $0.289$ & $0.655$ & F & F \\
 &  & 5 & $\mathbf{0.134}$ & $0.14$ & F & F & $\mathbf{0.134}$ & $0.143$ & $0.14$ & $0.155$ & $0.201$ & $0.251$ & $0.505$ & F & F \\
 &  & 10 & $\mathbf{0.133}$ & $0.14$ & F & F & $\mathbf{0.133}$ & $\mathbf{0.133}$ & $\mathbf{0.133}$ & $\mathbf{0.133}$ & $0.145$ & $0.259$ & $0.479$ & F & F \\
 &  & 20 & $0.133$ & $0.14$ & F & F & $0.133$ & $\mathbf{0.132}$ & $0.133$ & $0.134$ & $0.156$ & $0.212$ & $0.348$ & F & F \\
\addlinespace[2pt]
ROES & $x_1$ & 3 & $24.8$ & F & F & F & $8.63$ & $8.65$ & $10.5$ & $6.62$ & $5.18$ & $\mathbf{5.17}$ & F & F & F \\
 &  & 5 & $7.93$ & F & F & F & $5.55$ & $5.4$ & $5.51$ & $4.52$ & $\mathbf{3.92}$ & $5.2$ & F & F & F \\
 &  & 10 & $4.53$ & $13.1$ & F & F & $4.1$ & $\mathbf{3.81}$ & $3.83$ & $3.82$ & $3.85$ & $5.05$ & F & F & F \\
 &  & 20 & $3.59$ & $8.97$ & F & F & $3.56$ & $\mathbf{3.48}$ & $3.49$ & $3.55$ & $3.69$ & $4.61$ & F & F & F \\
\midrule
\multicolumn{3}{l}{Completed} & 43/44 & 39/44 & 18/44 & 9/44 & 44/44 & 44/44 & 44/44 & 44/44 & 42/44 & 40/44 & 36/44 & 30/44 & 24/44 \\
\bottomrule
\end{tabular}
}
\end{table*}

\newpage

\section{Detailed Evaluation Results}
\label{app:detailed-results}

Detailed evaluation results on each benchmark and order $\Zorder$ are given in \cref{tab:metric} for our evaluation metric~\eqref{eq:eval-metric}, and in \cref{tab:timing} for the computation time.

\begin{table*}[t]
\centering\small\setlength{\tabcolsep}{3pt}
\caption{Median reachability times in seconds over three measured repetitions, for the configurations and columns of \cref{tab:metric}; bold marks the fastest completed run of a row at the precision shown. Diagnostics and trace extraction are excluded.}
\label{tab:timing}
\begin{tabular}{lrrrrrrrrrrrrrr}
\toprule
 &  &  &  &  &  &  & \multicolumn{8}{c}{RAZeR, threshold $\tau$} \\
\cmidrule(lr){8-15}
Scenario & $\rho$ & Girard & Combastel & PCA & Scott & RAZeR & $0.05$ & $0.1$ & $0.2$ & $0.5$ & $1$ & $2$ & $5$ & $10$ \\
\midrule
LALO-W05 & 3 & $\mathbf{38.4}$ & F & F & F & $39.5$ & $47.2$ & $43.1$ & $48.9$ & F & F & F & F & F \\
 & 5 & $38.6$ & $\mathbf{35.2}$ & F & F & $35.7$ & $43.5$ & $43.7$ & $76.1$ & F & F & F & F & F \\
 & 10 & $\mathbf{34.3}$ & $\mathbf{34.3}$ & F & F & $36.9$ & $43.6$ & $43.6$ & $77.7$ & $79.5$ & F & F & F & F \\
 & 20 & $42.9$ & $\mathbf{34.2}$ & F & F & $37.6$ & $46.2$ & $46$ & $42.8$ & $42.7$ & F & F & F & F \\
\addlinespace[2pt]
LALO-W01 & 3 & $41.7$ & $\mathbf{33.7}$ & F & F & $36.4$ & $43.7$ & $42.5$ & $40.6$ & $40.2$ & $40.3$ & $40.1$ & $50$ & F \\
 & 5 & $39.4$ & $\mathbf{33.6}$ & F & F & $36.3$ & $46.3$ & $46.9$ & $40.5$ & $40.8$ & $40.7$ & $40.8$ & $45.5$ & F \\
 & 10 & $39.3$ & $\mathbf{33.9}$ & F & F & $36.8$ & $47.8$ & $44.7$ & $41$ & $41.4$ & $42$ & $41.5$ & $48.1$ & F \\
 & 20 & $36.5$ & $\mathbf{34.1}$ & F & F & $37.7$ & $47.8$ & $46$ & $43.3$ & $43.2$ & $43$ & $43.2$ & $49.4$ & $48.9$ \\
\addlinespace[2pt]
ROBE & 3 & $66$ & $\mathbf{65.1}$ & $66.6$ & F & $67.3$ & $81.1$ & $74.7$ & $72.6$ & $73.5$ & $74.1$ & $74.5$ & $84.6$ & $83.4$ \\
 & 5 & $65.2$ & $\mathbf{64.7}$ & $72.4$ & $70.9$ & $68.2$ & $75.6$ & $76.6$ & $87.5$ & $75.1$ & $75.4$ & $75.6$ & $86$ & $84.5$ \\
 & 10 & $66.7$ & $\mathbf{64.8}$ & $68.2$ & $74.6$ & $68.9$ & $79.1$ & $84$ & $76.3$ & $78$ & $78.6$ & $77.9$ & $88.8$ & $87.2$ \\
 & 20 & $67.5$ & $\mathbf{64.6}$ & $68.1$ & $78.2$ & $69.1$ & $87.3$ & $84.3$ & $81.6$ & $84$ & $84$ & $84.6$ & $95.1$ & $93.2$ \\
\addlinespace[2pt]
PRDE & 3 & $78.8$ & $\mathbf{75.3}$ & F & F & $79.3$ & $90.3$ & $88.5$ & $85$ & $87.4$ & $87.5$ & $91.8$ & $99.3$ & F \\
 & 5 & $77.8$ & $\mathbf{75.8}$ & F & F & $78.7$ & $85.9$ & $90.3$ & $89.3$ & $86.8$ & $222$ & $223$ & $98.9$ & $97.1$ \\
 & 10 & $\mathbf{74.3}$ & $74.9$ & $80.7$ & F & $79.3$ & $91.9$ & $91.4$ & $232$ & $244$ & $245$ & $240$ & $100$ & $102$ \\
 & 20 & $\mathbf{71.3}$ & $75.2$ & $73.6$ & F & $79.1$ & $98.9$ & $98.6$ & $256$ & $253$ & $253$ & $253$ & $102$ & $102$ \\
\addlinespace[2pt]
BIO & 3 & $\mathbf{7.58}$ & $8.03$ & F & F & $8.11$ & $9.32$ & $9.27$ & $23.4$ & $25$ & $26.1$ & $9.6$ & F & F \\
 & 5 & $\mathbf{7.37}$ & $7.84$ & F & F & $8.44$ & $9.04$ & $9.15$ & $9.8$ & $9.24$ & $9.37$ & $9.39$ & F & F \\
 & 10 & $\mathbf{7.3}$ & $7.84$ & F & F & $8.31$ & $9.34$ & $9.42$ & $9.42$ & $9.65$ & $9.91$ & $9.54$ & $11.1$ & F \\
 & 20 & $\mathbf{7.31}$ & $7.93$ & F & F & $8.42$ & $9.33$ & $9.48$ & $9.52$ & $10$ & $10.3$ & $10.5$ & $11.6$ & F \\
\addlinespace[2pt]
BRUS & 3 & $\mathbf{14.3}$ & $15.5$ & $14.6$ & F & $16.3$ & $17.5$ & $17.6$ & $17.3$ & $17$ & $17.1$ & $19.2$ & $20.5$ & $19.9$ \\
 & 5 & $\mathbf{14.5}$ & $15.2$ & $14.8$ & F & $16.3$ & $17.6$ & $17.7$ & $18.2$ & $17.6$ & $17.6$ & $18$ & $20$ & $19.9$ \\
 & 10 & $\mathbf{14.7}$ & $15.4$ & $14.9$ & F & $16.3$ & $17.7$ & $17.7$ & $18.4$ & $18.8$ & $18.3$ & $19$ & $20.2$ & $20.3$ \\
 & 20 & $\mathbf{14.6}$ & $15.3$ & $14.9$ & F & $16.5$ & $17.8$ & $17.8$ & $18$ & $18.1$ & $18.8$ & $18.5$ & $20.4$ & $20.6$ \\
\addlinespace[2pt]
GENE & 3 & $\mathbf{4.03}$ & $4.26$ & F & F & $4.43$ & $4.96$ & $5.05$ & $5.1$ & $5.72$ & $5.51$ & $5.29$ & $6.03$ & $5.96$ \\
 & 5 & $\mathbf{4.2}$ & $4.21$ & F & F & $4.66$ & $5.11$ & $5.09$ & $5.27$ & $5.36$ & $5.35$ & $6$ & $6.24$ & $6.12$ \\
 & 10 & $4.23$ & $\mathbf{4.18}$ & $4.87$ & $6.19$ & $4.63$ & $5.23$ & $5.17$ & $5.31$ & $5.45$ & $5.85$ & $5.88$ & $6.41$ & $6.45$ \\
 & 20 & $\mathbf{4.29}$ & $4.42$ & $4.44$ & $6.88$ & $4.83$ & $5.44$ & $5.36$ & $5.63$ & $5.62$ & $5.67$ & $6.08$ & $6.85$ & $6.91$ \\
\addlinespace[2pt]
JET & 3 & F & F & $\mathbf{23}$ & $24.2$ & $25.4$ & $27.7$ & $27.7$ & $27.8$ & $27.4$ & $27.4$ & $27.7$ & $31.6$ & $31.3$ \\
 & 5 & $\mathbf{22.6}$ & F & $23$ & $24.1$ & $24.9$ & $27.5$ & $27.7$ & $30.8$ & $30.5$ & $30.2$ & $29.5$ & $34$ & $31$ \\
 & 10 & $\mathbf{22.7}$ & $23.6$ & $23.1$ & $24.6$ & $25.1$ & $27.6$ & $28$ & $28$ & $28.1$ & $29.1$ & $73$ & $35$ & $31.5$ \\
 & 20 & $\mathbf{22.6}$ & $23.6$ & $23.3$ & $25.4$ & $25.1$ & $28.2$ & $28.5$ & $73.9$ & $77.5$ & $76$ & $76.6$ & $37.1$ & $33.2$ \\
\addlinespace[2pt]
LORE & 3 & $\mathbf{29.8}$ & $31$ & $30.6$ & F & $33.1$ & $36.2$ & $35.7$ & $93.4$ & $93$ & $93.2$ & $94.4$ & $50$ & $40.9$ \\
 & 5 & $\mathbf{30}$ & $30.9$ & F & F & $33.1$ & $35.9$ & $36.3$ & $93.6$ & $94.1$ & $35.2$ & $35.4$ & $48.9$ & $40.9$ \\
 & 10 & $\mathbf{30.1}$ & $31.2$ & F & F & $33.3$ & $36.6$ & $36.3$ & $35.2$ & $35.1$ & $35.4$ & $35.8$ & $53.9$ & $45.2$ \\
 & 20 & $\mathbf{30.1}$ & $30.9$ & $30.9$ & F & $33.3$ & $37$ & $37.4$ & $35.7$ & $36.2$ & $36.1$ & $36.5$ & $59.9$ & $46.8$ \\
\addlinespace[2pt]
LOVO & 3 & $\mathbf{19.7}$ & $20$ & F & F & $21.2$ & $23$ & $23$ & $22.1$ & $22.1$ & $22.2$ & $22.3$ & F & F \\
 & 5 & $\mathbf{19.3}$ & $20.1$ & F & F & $21.2$ & $23.3$ & $23.4$ & $22.2$ & $22.2$ & $22.2$ & $22.5$ & F & F \\
 & 10 & $\mathbf{19.4}$ & $20$ & F & F & $21.4$ & $23$ & $25.7$ & $22.4$ & $22.3$ & $22.6$ & $23$ & F & F \\
 & 20 & $\mathbf{19.5}$ & $19.9$ & F & F & $21.7$ & $24.6$ & $22.8$ & $22.7$ & $23$ & $23.1$ & $23.7$ & F & F \\
\addlinespace[2pt]
ROES & 3 & $\mathbf{17.8}$ & F & F & F & $19.7$ & $22.3$ & $21.5$ & $21$ & $20.9$ & $21.1$ & F & F & F \\
 & 5 & $\mathbf{17.6}$ & F & F & F & $19.5$ & $22.9$ & $21.7$ & $21$ & $21.2$ & $21.4$ & F & F & F \\
 & 10 & $\mathbf{17.6}$ & $18.3$ & F & F & $19.5$ & $22.6$ & $23.9$ & $21.1$ & $21.5$ & $21.7$ & F & F & F \\
 & 20 & $\mathbf{17.7}$ & $18.4$ & F & F & $19.7$ & $22.5$ & $25.1$ & $21.5$ & $22$ & $23$ & F & F & F \\
\midrule
\multicolumn{2}{l}{Completed} & 43/44 & 39/44 & 18/44 & 9/44 & 44/44 & 44/44 & 44/44 & 44/44 & 42/44 & 40/44 & 36/44 & 30/44 & 24/44 \\
\bottomrule
\end{tabular}

\end{table*}

\end{document}